\documentclass[journal,draftcls,onecolumn,12pt]{IEEEtranTCOM}
\usepackage{epsfig}
\usepackage{graphicx}
\usepackage{psfrag}
\usepackage{ifpdf}
\usepackage{cite}
\usepackage{siunitx}
\usepackage{hyperref}
\usepackage[utf8x]{inputenc}
\usepackage[top=1in, bottom=1in]{geometry}
\usepackage{fancyhdr}
\usepackage{lastpage}
\usepackage{oldstyle}
\usepackage{textcomp}
\usepackage{color}
\usepackage{placeins}
\usepackage{float}
\usepackage{tabularx,colortbl}
\graphicspath{{figures/Appendix/}}
\usepackage{amssymb} 
\usepackage{amsmath} 
\usepackage{subfig}
\usepackage{amsmath,epsfig}
\usepackage{amsfonts}
\usepackage{algpseudocode}
\usepackage{algorithm}
\usepackage{graphicx}
\usepackage{float}
\usepackage{epstopdf}
\usepackage{amsthm}
\usepackage{blindtext, graphicx}
\usepackage{amsmath}
\usepackage{fixltx2e}
\usepackage{textcomp}
\usepackage{graphicx}
\usepackage{epstopdf}
\usepackage{filecontents}
\usepackage{MnSymbol}%
\usepackage{wasysym}%
\usepackage{subfig}

\begin{document}

\begin{center}
 \LARGE An Overview of Physical Layer Security with Finite-Alphabet Signaling \\[10pt]
 \normalsize\today
\end{center}

\begin{center}
	\normalsize Sina Rezaei Aghdam*, Alireza Nooraiepour** ~and~ Tolga M. Duman*\\
	{\small (*) Dept. of Electrical and Electronic Engineering, Bilkent University,  Ankara, Turkey}\\
	{\small (**) WINLAB, Department of Electrical and Computer Engineering, Rutgers University, NJ, USA}\\
	{\small Email: \{aghdam,~duman\}@ee.bilkent.edu.tr, alireza.nooraiepour@rutgers.edu}\\
\end{center}


\begin{abstract}

Providing secure communications over the physical layer with the objective of achieving perfect secrecy without requiring a secret key has been receiving growing attention within the past decade. The vast majority of the existing studies in the area of physical layer security focus exclusively on the scenarios where the channel inputs are Gaussian distributed. However, in practice, the signals employed for transmission are drawn from discrete signal constellations such as phase shift keying and quadrature amplitude modulation. Hence, understanding the impact of the finite-alphabet input constraints and designing secure transmission schemes under this assumption is a mandatory step towards a practical implementation of physical layer security. With this motivation, this article reviews recent developments on physical layer security with finite-alphabet inputs. We explore transmit signal design algorithms for single-antenna as well as multi-antenna wiretap channels under different assumptions on the channel state information at the transmitter. Moreover, we present a review of the recent results on secure transmission with discrete signaling for various scenarios  including multi-carrier transmission systems, broadcast channels with confidential messages, cognitive multiple access and relay networks. Throughout the article, we stress the important behavioral differences of discrete versus Gaussian inputs in the context of the physical layer security. We also present an overview of practical code construction over Gaussian and fading wiretap channels, and we discuss some open problems and directions for future research.

\end{abstract}

\section{Introduction}
\label{sec:Introduction}
Wireless communication technologies have become an indispensable part of our everyday lives. As more and more data is being transmitted over wireless links, along with the reliability of the transmission, ensuring security of information transfer has become a challenging issue. Maintaining confidentiality of the transmitted data, preventing corruption of the transmitted information and verifying authenticity of communication are the most important security requirements in a wireless network. Traditionally, protecting confidentiality of data transmission is addressed via computation-based mechanisms such as encryption.
On the other hand, the security provided by these methods mainly relies on the conjecture that the encryption function is difficult to invert, which is not provable from a mathematical point of view. With the ever-increasing computing power, encryption may no longer prevent information leakage to sophisticated adversaries. Moreover, implementation of key-based secure communications requires complicated protocols for key distribution and management, which is highly challenging to implement in the cases of decentralized wireless networks.

Securing transmission at the physical layer is an alternative or a complement to the cryptographic solutions. The basic idea is to exploit the inherent randomness of the channel for achieving secrecy. Dissimilar to encryption based methods, in the schemes employing physical layer security, no constraint is placed on the computational capability of the eavesdropper. Furthermore, no key distribution/management is required in its implementation.

During the past decade there has been an extensive growth of interest in studying the capabilities of physical layer for securing communication. Several surveys and overview papers are available describing the state-of-the-art on this topic. For instance, fundamentals of physical layer security are comprehensively explained in \cite{Liang}-\hspace*{-3px}\cite{Shiu11}. A review of recent results ranging from point-to-point communications to multiuser scenarios is given in \cite{Mukherjee}. The authors in \cite{Hyadi} present a review of the recent research results on physical layer security under channel state information (CSI) uncertainty in different nodes. A comprehensive overview on various multiple antenna techniques in physical layer security is provided in \cite{Chen17}. The potentials of physical layer security for safeguarding data confidentiality in the fifth generation (5G) network is studied in \cite{NYang15} and two updated summaries of the recent research results on physical layer security techniques for 5G and next generation wireless networks are given in \cite{Wu_phy_sec} and \cite{YLiu_survey}.

The existing literature reviews provide a comprehensive understanding of the fundamentals of physical layer security. However, less is said about the practical aspects. Adoption of physical layer security techniques in the next generation networks requires a number of important practical issues to be addressed. With this motivation, a high-level review on these issues is provided in a recent IEEE Communications Magazine paper \cite{Trappe}. While \cite{Trappe} provides the big picture, our objective in this survey is to shed light on two important (and highly related) practical aspects, namely, secure transmission with discrete signaling and practical code construction. We present a detailed review of the related recent results, underline the important behavioral differences between the Gaussian inputs and discrete signaling, and study the transmit signal design problem for securing communications in a variety of scenarios. We also provide a review of recent results on practical code design for secrecy, and make connections with the information theoretic results with finite inputs.

\begin{figure}[h!]
	\begin{center}
		\begin{tabular}{c}
			\mbox{\psfig{figure=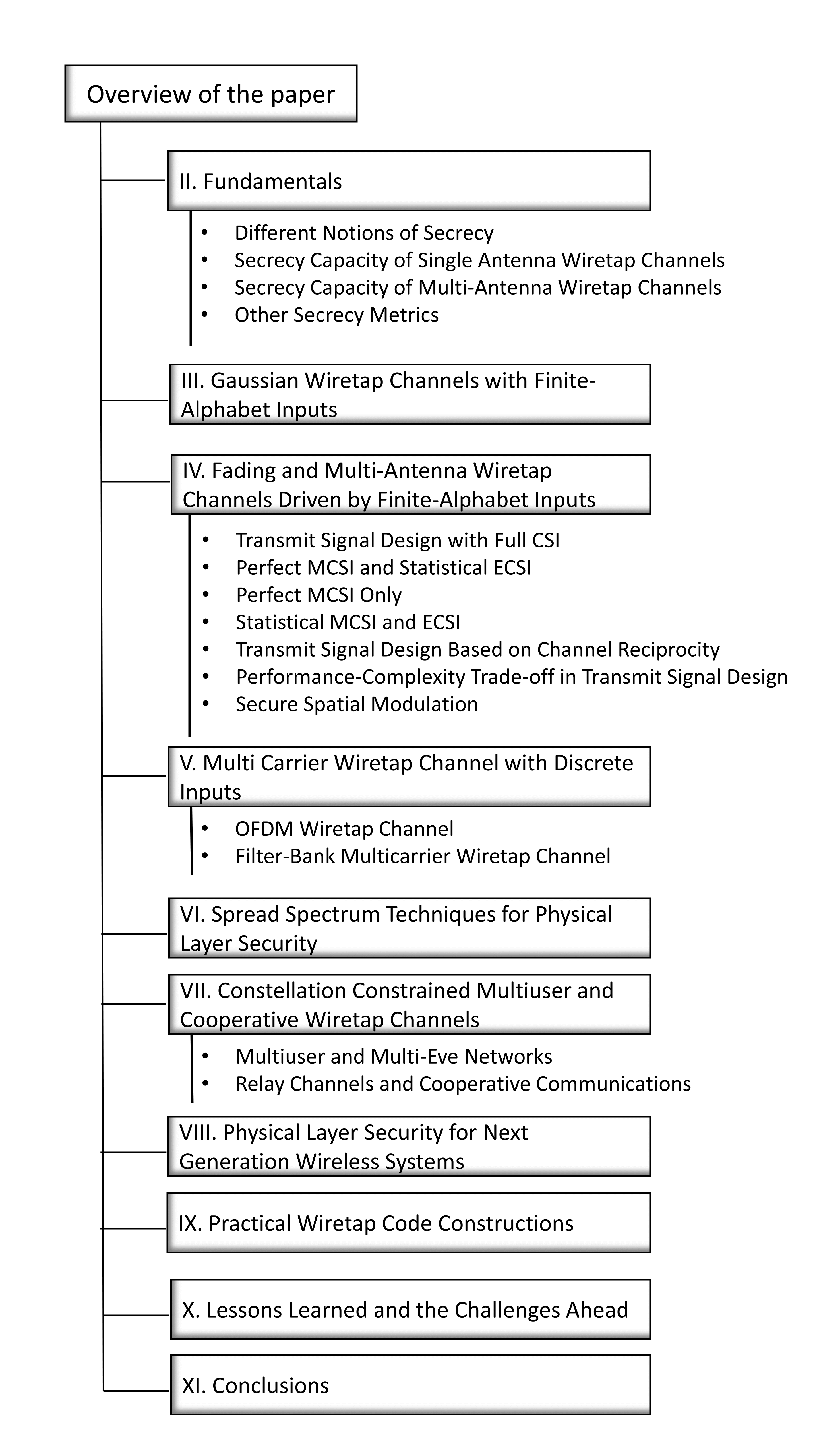, width=7.75cm}}
		\end{tabular}
	\end{center}
	\caption{ The organizational structure of the paper.} \label{Structural}
\end{figure}

The remainder of the paper is organized as follows. In Section \ref{sec:Fund}, we provide a description of the fundamental concepts in physical layer security including the wiretap channel model and different secrecy metrics. In Section \ref{sec:p2pSISO}, we discuss the differences between the secrecy performance of constellation-constrained and Gaussian inputs over Gaussian wiretap channels. An overview of secure multi-antenna transmission schemes with finite-alphabet inputs is provided in Section \ref{sec:p2pMIMO}. Sections \ref{sec:multicarrier}, \ref{sec:SS} and \ref{sec:multiuser} discuss the scenarios with multi-carrier transmission, spread spectrum communications and multiuser channels, respectively. Section \ref{sec:5G} focuses on applications of physical layer security techniques in the next generation wireless networks including 5G. An overview of the practical code design approaches for secrecy is presented in \ref{sec:Coding}, followed by concluding remarks and future research directions in Section \ref{sec:challenges}. The organizational structure of this paper is given in Fig. \ref{Structural}.

\textit{Notation}: Vectors and matrices are denoted with lowercase and uppercase bold letters. Non-bold letters are used to denote scalar values and calligraphic letters denote sets. The expectation and probability mass (or density) function of a random variable $W$ is denoted by $\mathbb{E}_{W}\{.\}$ and $P_W(.)$, respectively. $Pr(.)$ is used to denote the probability of an event. $A^n$ stands for a sequence of length $n$, i.e., $A^n = \{A(1),A(2),\ldots,A(n)\}$. The notation $\mathcal{CN}(\textbf{m},\textbf{R})$ denotes a circularly symmetric complex Gaussian random vector with mean vector $\textbf{m}$ and covariance matrix $\textbf{R}$.

 
\section{Fundamentals}
\label{sec:Fund}

The wiretap channel model, originally introduced by Wyner in \cite{Wyn74}, is the most basic scenario in the study of physical layer security. In this model, while the sender Alice wishes to transmit a message signal to a legitimate receiver Bob; a third party, Eve, is present with the capability of eavesdropping on Alice's signal. The channel between Alice and Bob, and the channel between Alice and Eve are referred to as the main channel and the eavesdropper's channel, respectively. In physical layer security, one is interested in transmitting in such a way to maximize the transmission rate over the main channel while keeping the eavesdropper ignorant about the message. This leads to the notion of secrecy capacity, which is defined as the maximum rate at which the transmitter can use the main link so as to deliver its message to the legitimate receiver in a way that the eavesdropper cannot successfully obtain any information about it.
Characterization of the secrecy capacity is one of the most fundamental problems in the literature of physical layer security in that it can provide us with vital implications on how secure transmission techniques can be designed.

\subsection{Different Notions of Secrecy}
Consider a transmitter who wishes to transmit a message $M$ to a legitimate receiver while trying to keep an eavesdropper ignorant about it. $M$ is mapped to a codeword $X^n$ using a stochastic encoder with $n$ denoting the number of channel uses. Then, $X^n$ is transmitted, and $Y^n$ and $Z^n$ are received at the legitimate receiver and the eavesdropper, respectively. The eavesdropper's level of uncertainty is quantified using the equivocation rate which is given by
\begin{equation}
R_e = \frac{1}{n} H(M|Z^n),
\end{equation}
which is nothing but a measure of how unlikely it is that the eavesdropper can infer source information from its received signal. According to this definition, large equivocation rates are equivalent to higher secrecy levels.

There are different metrics for measuring the secrecy guaranteed by a specific scheme among which the information-theoretic ones are recognized as the strongest. Perfect secrecy is achievable in a system if the message $M$ and its corresponding encoder output $X^n$ are statistically independent \cite{Shannon}. This can be expressed as $I(M;X^n) = 0$, which means that the mutual information between the message and the encoded signal is exactly zero. This is to say, the eavesdropper is not capable of recovering the message via observation of the encoder output in view of the fact that this encoded signal does not provide any information about it. For this form of secrecy to be guaranteed, a secret key of the length at least equal to the length of the message should be used. However, weaker requirements can be adopted for evaluation of the secrecy in practical scenarios. One example of such notions is semantic secrecy, which is achieved if it is guaranteed that $Z^n$ does not increase the probability of guessing an arbitrary function $f$ of $M$ (with an arbitrary distribution $P_M$) by more than $\epsilon > 0$ if $Adv(M;Z^n)<\epsilon$, where advantage $Adv(M;Z^n)$ is defined as
\begin{equation}
Adv(M;Z^n)=\max\limits_{f, P_M}\big(\sum_{z^n}P_{Z^n}(z^n)\max\limits_{f_i\in supp(f)} Pr(f(M)=f_i|z^n)-\max\limits_{f_i\in supp(f)}Pr(f(M)=f_i)\big),
\end{equation}
where $supp(.)$ stands for the support of the function in its argument. 

Assuming a uniform distribution for the message $M$, a system is called to operate with strong secrecy if $ \lim_{n\to \infty} I(M;Z^n)=0$, which means that if we consider codewords of $n$ symbols being transmitted by Alice as $X^n$, the information leaked by observation of the received vector $Z^n$ shall go to zero as $n$ goes to infinity. On the other hand, the condition $\lim_{n\to \infty} \frac{1}{n} I(M;Z^n)=0$ is referred to as weak secrecy. In particular, weak secrecy requires the asymptotic rate of information leakage to be sublinear in $n$. Different notions of secrecy, which are used to measure the secrecy guaranteed by a specific scheme are summarized in Table~\ref{SecMet}.

\begin{table}[h!]
	\caption{Four notions of secrecy.}\label{SecMet}
	\begin{center}
		\begin{tabular}{ | l | l  | l  |}
			\hline
			Notion & Requirement\\ \hline
			Perfect Secrecy \cite{Shannon} & $I(M;X^n)=0$ \\ \hline
			Semantic Secrecy \cite{semantic} & $Adv(M;Z^n)<\epsilon$ \\ \hline
			Strong Secrecy \cite{Maurer} & $ \lim_{n\to \infty} I(M;Z^n)=0$ \\ \hline
			Weak Secrecy \cite{Wyn74} & $\lim_{n\to \infty} \frac{1}{n} I(M;Z^n)=0$ \\ \hline
		\end{tabular}
	\end{center}
\end{table}   

\subsection{Secrecy Capacity of Single-Antenna Wiretap Channels}
The notion of secrecy capacity was originally introduced by Wyner in \cite{Wyn74} for degraded wiretap channels. In the scenario studied in \cite{Wyn74}, the main channel and the eavesdropper's channel are assumed to be  discrete memoryless channels (DMCs) where Eve's observation of the transmitted signal is a degraded version of the signal received by Bob. Wyner showed that, the secrecy capacity for this scenario can be expressed as the difference of the mutual information between Alice and Bob with that of Alice and Eve, maximized over all input distributions. An extension of the notion of secrecy capacity to the case of additive white Gaussian noise (AWGN) channels has been accomplished in \cite{Leung}. It is shown that, for the case of a Gaussian wiretap channel, nonzero secrecy capacity can be attained if Bob receives the signal through a less noisy channel.

While the scenarios considered in \cite{Wyn74} and \cite{Leung} put the eavesdropper at a disadvantage, Csisz\'ar and K\"orner have studied the problem of secrecy for non-degraded channels in \cite{Csiszar}. In this channel model, referred to as broadcast channel with confidential messages, the message transmitted by Alice contains a common message intended to both Bob and Eve as well as a secret message, which is intended for Bob only, and is needed to be kept secret from Eve. 
It is shown that the secrecy capacity over a discrete memoryless wiretap channel (DMWC) is given by
\begin{equation}\label{SecCapacity}
C_s = \max_{P_{X|U}, P_U}{I(U;Y) - I(U;Z)}, 
\end{equation}
\noindent where $U$ is an auxiliary random variable. With a given $P_{YZ|X}(y,z|x)$, the secrecy capacity is achieved by maximizing the difference of the mutual information terms corresponding to the main and the eavesdropper's channels over all joint distributions $P_{UX}(u,x)$ where $U$ satisfies the Markov chain relationship $U \rightarrow X \rightarrow YZ$.

In order to study the physical layer security in wireless communication scenarios, we need to extend the wiretap channel to a model that takes into account the fading phenomenon. Assuming a narrowband transmission, fading is modeled as  multiplicative gains over the main and eavesdropper's channels, denoted by $H_B$ and $H_E$, respectively. The first characterization of the role of fading in providing physical layer security is given in \cite{Barros06} where quasi-static fading channels are considered towards the legitimate receiver and the eavesdropper. It is shown via analyzing the secrecy outage probability that, in presence of fading, secure transmission is possible even in the scenarios where the eavesdropper has a better SNR. The authors in \cite{Liang_BCC} obtain the ergodic secrecy capacity region for the fading broadcast channel with confidential messages (BCC) by deriving the boundary achieving optimal power allocation strategies. The secrecy capacity over block fading channels is derived in \cite{Gopala} under the assumption that each coherence interval is long enough to allow for invoking proper random coding arguments. Given that the transmitter knows either the main CSI (MCSI) or both the main and the eavesdropper CSI (ECSI) perfectly, an achievable secrecy rate can be formulated as \cite{Gopala}:
\begin{equation}\label{RsFading_ab}
R_{s} = \mathbb{E}_{H_B,  H_E} \bigg\{\big[I(X;Y|H_B) - I(X;Z|H_E) \big]^{+}\bigg\},
\end{equation}
where $[a]^+ = \max\{a, 0\}$. It is shown in \cite{Gopala} that the secrecy capacity that is achieved under the full CSI assumption serves as an upper bound on the secrecy capacity when only MCSI is known at the transmitter. This is due to the fact that with the knowledge of the instantaneous ECSI, transmitter is capable of realizing a more efficient transmission, which provides additional gains in terms of secrecy rates. For instance, when perfect knowledge of both channels are available, the transmitter can optimize the transmit power according to the instantaneous values of the channel gains $H_B$ and $H_E$, providing a higher secrecy rate with respect to the case where the optimization of the transmit power is carried out according to the main channel only.

Under the assumption that the transmitter does not know the realizations of the channels, and it only has their statistics, assuming that each receiver knows its own channel, the secrecy capacity can be obtained with the aid of the results for the case of DMWC \cite{Csiszar} as 
\begin{equation}\label{SecCapa1}
C_s = \max_{P_{X|U}, P_U}{I(U;Y,H_B) - I(U;Z,H_E)}, 
\end{equation}
\noindent where $U$ satisfies the Markov chain relationship $U \rightarrow X \rightarrow (Y,H_B), (Z,H_E)$.  This is obtained by treating $H_B$ and $H_E$ as channel outputs at the legitimate receiver and at the eavesdropper, respectively \cite{Lin16}.  Using the chain rule of mutual information and noting the fact that $H_B$ and $H_E$ are independent of $X$ and $U$, the expression in (\ref{SecCapa1}) can be modified as
\begin{equation}
C_s  = \max_{P_{X|U}, P_U} \mathbb{E}_{H_B} I(U;Y|H_B) - \mathbb{E}_{{H}_E} I(U;Z|H_E) \label{SecCap2}.
\end{equation} 
Determining the optimal joint distribution $P_{UX}(u,x)$ and the resulting exact secrecy capacity is an open problem, however, by ignoring the channel prefixing, the performance can be quantified using an achievable ergodic secrecy rate as given by
\begin{equation}\label{RsFading_c}
R_{s} = \big[\mathbb{E}_{H_B} I(X;Y|H_B) - \mathbb{E}_{H_E} I(X;Z|H_E) \big]^{+}.
\end{equation}

\subsection{Secrecy Capacity of Multi-Antenna Wiretap Channels}

Multiple-input multiple-output (MIMO) wiretap channel is an extension of the wiretap channel to a scenario where all the nodes, namely, Alice, Bob and Eve, are equipped with multiple antennas. Under the full CSI assumption, secrecy capacity for this setup is determined independently by Khisti and Wornell \cite{Khisti}, and Oggier and Hassibi \cite{Hassibi}. The results in \cite{Csiszar} are employed by Khisti and Wornell so as to develop a genie-aided upper bound on the secrecy capacity in the general case of non-degraded MIMO wiretap channels. While their characterization is through a saddlepoint, Oggier and Hassibi propose an alternative approach through a single optimization process. It is shown in both \cite{Khisti} and \cite{Hassibi} that the achievable secrecy rate is maximized when the channel input is Gaussian, and the secrecy capacity of a MIMO wiretap channel is given by
\begin{equation}
C_s = \max_{\textbf{K}_{\textbf{x}} \succeq 0, Tr(\textbf{K}_{\textbf{x}})=P} \log \det(\textbf{I} + \textbf{H}_B \textbf{K}_{\textbf{x}} \textbf{H}_B^H) - \log \det(\textbf{I} + \textbf{H}_E \textbf{K}_{\textbf{x}} \textbf{H}_E^H),  
\end{equation}
where $\textbf{K}_{\textbf{x}} = \mathbb{E}\{\textbf{x}\textbf{x}^H\}$, and $\textbf{H}_B \in \mathbb{C}^{N_B \times N_t}$ and $\textbf{H}_E \in \mathbb{C}^{N_E \times N_t}$ are the channel matrices corresponding to the legitimate receiver and the eavesdropper, respectively.

An alternative characterization of the secrecy capacity of the MIMO wiretap channel is presented by Liu and Shamai in \cite{Shamai} where they consider a more general matrix constraint on the channel input. We highlight that, while \cite{Khisti}--\hspace{-0.5px}\cite{Shamai} prove that the optimal input is Gaussian, the optimal input covariance matrix needs to be determined using numerical optimization approaches (see, e.g., \cite{QLi}--\hspace{-0.5px}\cite{Loyka}). A set of equations describing the optimal channel input under partial CSI at the transmitter (CSIT) along with algorithms for obtaining the solution are given in \cite{JLi}.

\subsection{Other Secrecy Metrics}

Secrecy capacity is the most widely used measure for evaluation of the secrecy performance. However, a complete characterization of the secrecy capacity region is prohibitive in some scenarios. In such cases, alternative metrics can be adopted as surveyed below.

%
%

\subsubsection{Secrecy Outage Probability}
Secrecy outage occurs when the instantaneous secrecy capacity falls below a target secrecy rate $R_s^{\epsilon}$. In delay-critical applications and for scenarios with quasi-static fading, encoding over multiple channel states is not possible. In these cases, secure transmission schemes can be designed such that the probability of occurrence of an outage event is minimized. Obviously, this serves as a weaker notion than secrecy capacity. This is due to the fact that, with this formulation, secrecy is not guaranteed for the entire transmission duration.

\subsubsection{Error Probability Based Metrics}\label{Error Probability/SINR Based Metrics}
In a number of works on physical layer security, secure transmission schemes are designed based on constraining the bit error rate at the legitimate receiver and the eavesdropper to pre-specified threshold values. This leads to definition of a parameter called \textquotedblleft security gap," which quantifies the minimum required difference between the received SNR values at Bob and Eve to ensure that the bit error rate (BER) at Bob is smaller than a threshold while that at the Eve's receiver is larger than another threshold \cite{Klinc}. One may note that imposing a high BER at the eavesdropper does not satisfy any of the requirements given in Table \ref{SecMet}. However, security gap serves as a practical measure of secrecy in a number of applications such as explicit and implementable code design for physical layer secrecy. Some studies (e.g., \cite{Liao}) consider the difference in the signal-to-interference-plus-noise ratio (SINR) at the legitimate receiver and the eavesdropper as a secrecy metric. This is also a weak notion, which does not guarantee secrecy in an information theoretic sense. 



\section{Gaussian Wiretap Channels with Finite-Alphabet Inputs}
\label{sec:p2pSISO}

The secrecy capacity achieving input distribution over a Gaussian MIMO wiretap channel is proved to be Gaussian \cite{Leung}. However, since Gaussian signals take on a continuum of values, their detection complexity is considerably high. In addition, noting that the amplitude the Gaussian signals is unbounded, these signals have high peak to average power ratios. Accordingly, Gaussian signaling is not typically used in practice, and instead the transmission is carried out with the aid of discrete inputs drawn from standard constellations such as PSK or QAM. Hence, it is important to study the implications of discrete signaling in the context of physical layer security. 

\begin{figure}[t!]
	\begin{center}
		\begin{tabular}{c}
			\mbox{\psfig{figure=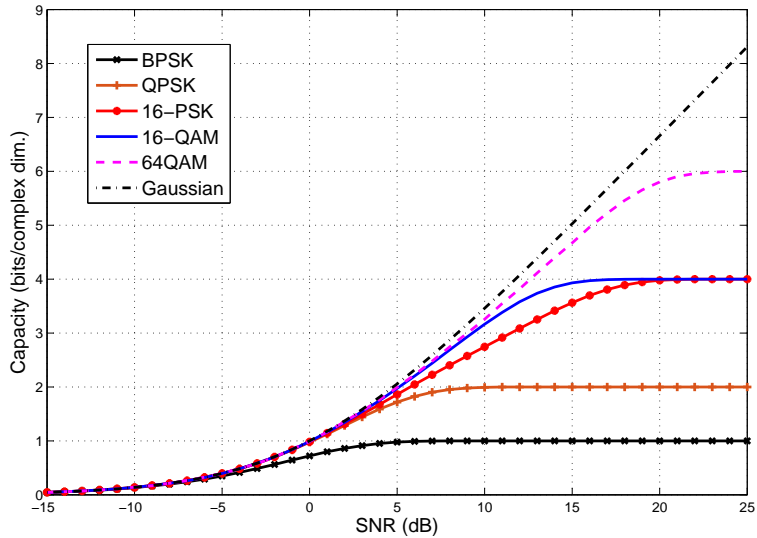,width=12cm}}
		\end{tabular}
	\end{center}
	\caption{Capacity of an AWGN channel with Gaussian and constellation-constrained inputs.}\label{AWGN_cap}
\end{figure}

\begin{figure}[t!]
	\begin{center}
		\begin{tabular}{c}
			\mbox{\psfig{figure=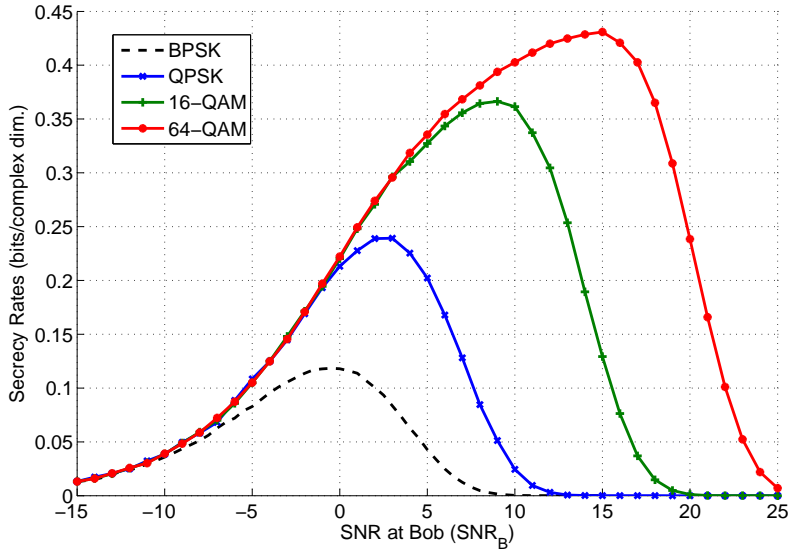,width=12cm}}
		\end{tabular}
	\end{center}
	\caption{Secrecy rates with PSK and QAM over a degraded Gaussian wiretap channel ($\text{SNR}_E = \text{SNR}_B - 1.5 ~\text{dB}$).}\label{diff_mod}
\end{figure}

As a first step in studying physical layer security under the finite-alphabet input assumption, the impacts of standard constellations on the achievable secrecy rates of Gaussian wiretap channels are studied in \cite{Raghava} and \cite{Rodrigues}. In \cite{Raghava}, the authors evaluate the constellation-constrained secrecy capacity and highlight an important behavioral difference between finite-alphabet (e.g., PSK and QAM) and Gaussian inputs, that is, for a fixed noise variance at Eve, the secrecy rate curves for PSK or QAM plotted against the SNR have global maxima at finite SNR values. Investigation of the secrecy capacity of the pulse amplitude modulation (PAM) inputs over a degraded Gaussian wiretap channel in \cite{Rodrigues} leads to a similar conclusion. In addition, the authors in \cite{Rodrigues} demonstrate that when finite constellations are employed, using all the available power for the information bearing signal transmission may not be optimal. The reason behind these observations has its roots in the mutual information expressions. That is, the capacity under an average power constraint over an																																																																																																																																																																																																																																																																																																																																																																																																																																																																																																																																																																																																																																																																																																																																																																																																																																																																																																																																																																																																																																																																																																																																																																																																																																																																																																																																																																																																											AWGN channel is given by
\begin{equation} \label{Capacity}
C = \log_2 (1 + SNR),
\end{equation}
in bits per complex dimension, and it is an increasing function of the SNR. On the other hand, the capacity of the constellation-constrained AWGN channel can be calculated as \cite{Biglieri}
\begin{equation} \label{CC_cap}
I(X;Y) = \log_2 |\mathcal{X}| - \mathbb{E}_{X,Y}\Bigg\{\log_2 \frac{\sum_{x' \in \mathcal{X}}^{}p_{Y|X'}(y|x')}{p_{Y|X}(y|x)} \Bigg\},
\end{equation}
where $\mathcal{X}$ is a constellation set and $|\mathcal{X}|$ denotes its cardinality. Dissimilar to the capacity expression in (\ref{Capacity}), the mutual information term in (\ref{CC_cap}) converges to $\log_2|\mathcal{X}|$ at high SNRs (see Fig. \ref{AWGN_cap}). This means that by increasing the transmit power, the mutual information terms $I(X;Y)$ and $I(X;Z)$ (calculated using (\ref{CC_cap})) will reach their saturation value, and the secrecy rate drops to zero. This is also verified in Fig. \ref{diff_mod}, where the secrecy rates are obtained by evaluating the difference of the mutual information terms corresponding to the main and eavesdropper's channels, under the assumption that the SNR at the eavesdropper (denoted by $\text{SNR}_e$) is lower than the legitimate receiver's SNR (denoted by $\text{SNR}_b$) by $1.5 ~dB$. This also clarifies that using all the available transmit power for information transmission in the high SNR regime is not optimal \cite{Rodrigues}. Furthermore, it can be observed from Fig. \ref{diff_mod} that higher secrecy rates are achieved over Gaussian wiretap channels by increasing the modulation orders. This is because by increasing the modulation order, roughly speaking, the distribution of the channel input can be made to resemble the optimal Gaussian signal more closely for a wider range of SNRs.

The impact of output quantization on the secrecy rate of binary-input Gaussian wiretap channels is investigated in \cite{ChaoQi}. It is shown through a theoretical analysis that when the legitimate receiver has unquantized inputs where the eavesdropper has (binary) quantized outputs higher secrecy rates are achievable with respect to the case where both receivers have binary quantized or unquantized outputs.


To conclude this section, we note that while it may be tempting to argue that, as in the case of transmission over an AWGN channel with no security considerations, one can design general transmission schemes using Gaussian codebooks and simply adopt the ideas to the case of inputs from standard constellations (e.g., PSK, QAM, etc.), this approach does not extend in a straightforward manner to the case of physical layer security. The optimal transmit signal design strategies under Gaussian input assumption lead to considerable losses when applied to inputs drawn from standard constellations. For instance, it is shown in \cite{Rodrigues} the optimal power allocation strategy with PAM inputs is in sharp contrast to that of Gaussian inputs. This motivates development of new algorithms and ideas for practical constellation-constrained channels to achieve security at the physical layer.

\section{Fading and Multi-Antenna Wiretap Channels Driven by Finite-Alphabet Inputs}
\label{sec:p2pMIMO}

Fading introduces new potentials for securing communications at the physical layer. Specifically, randomness due to channel fluctuations can be exploited opportunistically by a transmitter to guarantee secrecy even in the scenarios where the eavesdropper possesses a higher SNR (on average) than the legitimate receiver. Dissimilar to the Gaussian wiretap channel where Gaussian input is secrecy capacity achieving, in the presence of fading, discrete signaling may achieve higher secrecy rates. For instance, it is shown in \cite{ZLi} that, when Bob's channel gain is on average worse than that of Eve's, QAM inputs achieve higher secrecy rates at low and moderate SNR regimes. This is because the discrete nature of QAM inputs limits the leakage at the eavesdropper whose channel is unusually good. At sufficiently high SNRs, however, the secrecy rate with QAM inputs drops to zero similar to what is observed for the case of Gaussian wiretap channels.

\begin{figure}[t!]
	\begin{center}
		\begin{tabular}{c}
			\mbox{\psfig{figure=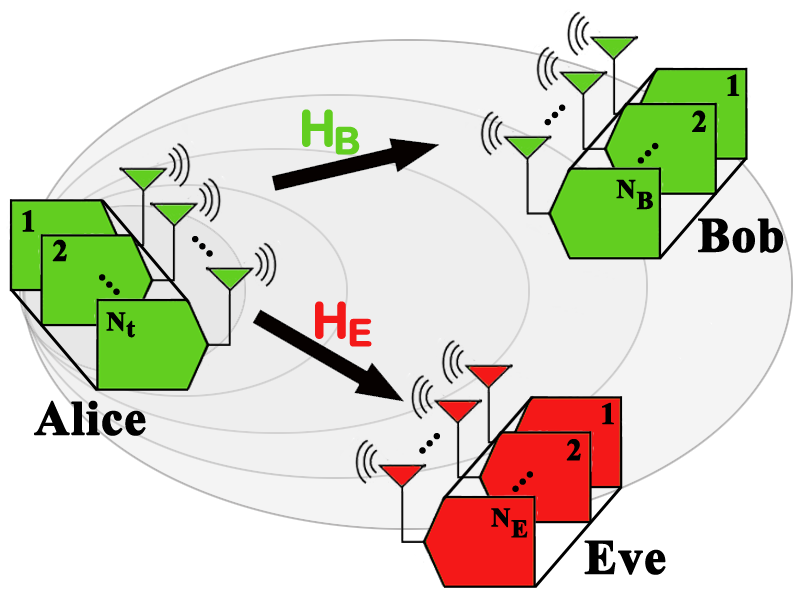, width=12cm}}
		\end{tabular}
	\end{center}
	\caption{MIMO wiretap channel.} \label{MIMOwiretap}
\end{figure}

Taking advantage of the spatial degrees of freedom introduced by multi-antenna transmission can serve as an efficient solution to prevent secrecy rate of standard constellations from dropping to zero at high SNRs. In the remainder of this section, we focus on a system model where Alice, Bob and Eve are equipped with $N_t$, $N_B$ and $N_E$ antennas, as depicted in Fig. \ref{MIMOwiretap}. The received signals are given by
\begin{eqnarray}
\textbf{y} = \textbf{H}_B \textbf{x} + \textbf{n}_y,\\
\textbf{z} = \textbf{H}_E \textbf{x} + \textbf{n}_z,
\end{eqnarray}
where $\textbf{H}_B$ and $\textbf{H}_E$ are the channel matrices corresponding to the legitimate receiver and the eavesdropper, respectively, while $\textbf{n}_y$ and $\textbf{n}_z$ denote circularly symmetric complex AWGN vectors, elements of which follow $\mathcal{CN}(0,\sigma{}^{2}_{\textbf{n}_y})$ and $\mathcal{CN}(0,\sigma{}^{2}_{\textbf{n}_z})$, respectively.

Transmit precoding and artificial noise injection serve as important approaches for enhancing secrecy \cite{Mukherjee}. Via precoding, it is possible to strengthen (or weaken) the transmitted signals in certain directions. Accordingly, it can be used as an effective tool to increase the quality difference between the signals received at the legitimate receiver and the eavesdropper. Beamforming is a special case of precoding where the transmitter is restricted to using rank-one signals. It has a lower complexity with respect to the general transmit precoding solution, achieved at the price of some performance loss. Furthermore, injection of artificial noise \cite{Goel} can degrade the reception at the eavesdropper effectively while having no (or minimal) effect on the signal received at the legitimate receiver. We also note that, to take advantage of the potentials offered by the channel fading, the transmitter needs a certain level of knowledge on the CSI of both channels. 

In what follows, we review the existing solutions for secure transmission over MIMO wiretap channels with finite-alphabet inputs under different CSIT assumptions.

\subsection{Transmit Signal Design with Full CSI}

The most optimistic assumption regarding the transmitter's CSI knowledge is availability of perfect instantaneous realizations of the channel matrices $\textbf{H}_B$ and $\textbf{H}_E$. This assumption is justifiable in the scenarios where Eve is an authorized user in the network, however, she should be kept ignorant about the confidential messages transmitted from Alice to Bob. In this case, the transmitter can take advantage of the full CSI knowledge to design a precoder, which results in the maximal quality difference between the signals received at the legitimate receiver and the eavesdropper.

With the aid of the perfect CSI corresponding to both channels, a generalized singular value decomposition (GSVD) precoding solution is proposed in \cite{Bashar} using which the MIMO multi-antenna eavesdropper (MIMOME) channel is converted into a bank of parallel channels, and a power allocation strategy is formulated to maximize the achievable secrecy rate.

\begin{figure}[t!]
	\begin{center}
		\begin{tabular}{c}
			\mbox{\psfig{figure=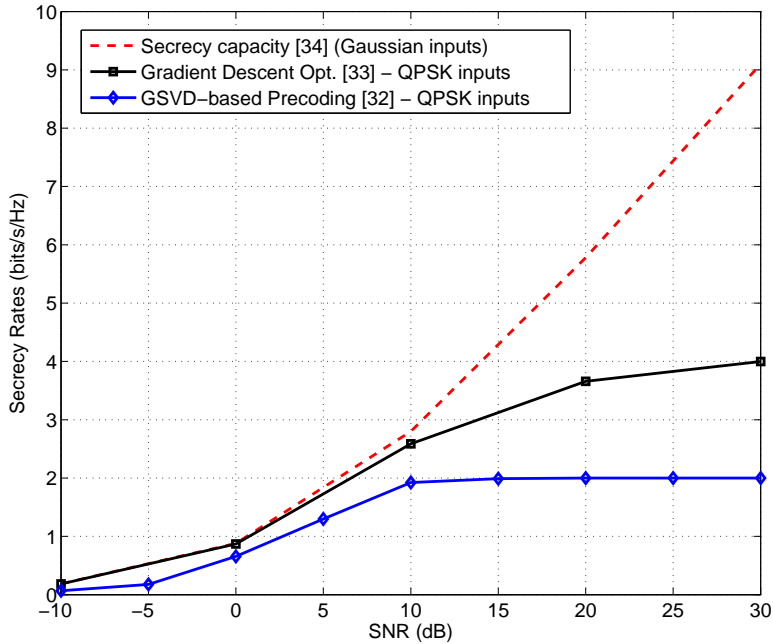,width=12cm}}
		\end{tabular}
	\end{center}
	\caption{Secrecy rates with QPSK inputs over fixed channels given in (\ref{channel_Bob}) and (\ref{channel_Eve}).}\label{fullCSI}
\end{figure}

The necessary conditions for optimum transmit precoding is derived in \cite{Wu}, and it is demonstrated that the GSVD precoding \cite{Bashar} is suboptimal. Alternatively, a gradient descent optimization is proposed in which the precoder matrices are updated with steps proportional to the gradient of the instantaneous secrecy rate. Furthermore, it is shown that transmission along the null-space of the eavesdropper's channel is the optimal secure transmission strategy at the high-SNR regime. This is because, with a precoder matrix along the null-space of the eavesdropper's channel, her achievable rate is suppressed to zero, and when the SNR is sufficiently high, the rate at the legitimate receiver approaches the maximum value that can be achieved with the use of specific finite-alphabet inputs, which guarantees that the secrecy rate is maximized. However, it should be noted that, when the number of transmit antennas is less than or equal to the number of antennas at the eavesdropper, it is not possible to suppress her information rate by transmitting along the null-space of $\textbf{H}_E$. Therefore, for maximizing secrecy rates, only part of the available power should be allocated to data transmission. Given $N_{t}>N_B$, the authors in \cite{Wu} suggest to use the excess power for transmission of an artificial noise signal along the null-space of $\textbf{H}_B$ so as to prevent the secrecy rates from dropping to zero. Fig. \ref{fullCSI} compares the achievable secrecy rates by the schemes proposed in \cite{Bashar} and \cite{Wu} for a wiretap channel with $N_t = 2$, $N_B = 1$ and $N_E = 1$, and fixed channels given by\footnote{This example has also been considered in \cite{Wu} and \cite{Khisti_MISOSE}.}
\begin{eqnarray} \label{channel_Bob}
\textbf{h}_B = \begin{bmatrix}
0.5128-0.3239j & -0.8903-0.0318j
\end{bmatrix},\\ \label{channel_Eve}
\textbf{h}_E = \begin{bmatrix}
0.3880+1.2024j & −0.9825+0.5914j
\end{bmatrix}.
\end{eqnarray}
The precoding approach proposed in \cite{Wu} outperforms the GSVD-based precoding \cite{Bashar}. This improved performance is attained at the price of increased computational complexity. Furthermore, comparing the results given in Fig. \ref{fullCSI} with those of Fig. \ref{diff_mod} for QPSK inputs over the Gaussian wiretap channel reveals the advantages offered by fading and CSI-based precoding approaches.

An important conclusion that can be drawn from the results in \cite{Bashar} and \cite{Wu} is that, as in the case of non-fading channels, the structure of the optimal precoders under the finite-alphabet input assumption is different from that of the precoders, which are optimal for Gaussian inputs. For example, over MISO multi-antenna eavesdropper (MISOME) channels, the optimal transmission strategy is beamforming along the eigenvector corresponding to the largest generalized eigenvalue of $(\textbf{H}_b, \textbf{H}_e)$ \cite{Khisti_MISOSE}. However, these precoders (which correspond to signaling with rank one covariance), when applied to finite-alphabet inputs, undergo a considerable loss with respect to the precoders specifically optimized for this case (see, e.g., Fig. 7 in \cite{Wu}).

As an alternative method to design a suboptimal precoder, one may also replace the secrecy capacity with practical metrics (as opposed to those ensuring secrecy in an information-theoretic sense) such as SINR-based metrics. For instance, the authors in \cite{Khandaker} design artificial noise beamformers, which serve as constructive interference to the legitimate receiver (improving Bob's SINR) while disrupting the reception at the eavesdropper (degrading Eve's SINR).
In Table \ref{tbl:precodingtechn}, we summarize the precoder design approaches with full CSI.

\begin{table}[h!]
	\centering
	\caption{Precoder design with full CSI.}
	\label{tbl:precodingtechn}
	\begin{tabular}{|l|p{1.9cm}|p{8.3cm}|}
		\hline
		{\scriptsize Authors} & {\scriptsize System Model} & {\scriptsize Contributions}                                                                                                                                   \\ \hline \hline
	{\scriptsize 	S. Bashar \textit{et al.} \cite{Bashar}} & {\scriptsize MIMOME}& {\scriptsize  Propose a GSVD-based precoding algorithm.}\\ \hline
	{\scriptsize 	Y. Wu \textit{et al.} \cite{Wu}} & {\scriptsize MIMOME} & {\scriptsize Derive the necessary conditions for optimality of the precoder matrix and develop a gradient descent based precoder optimization algorithm.}\\ \hline
	{\scriptsize 	M. R. A. Khandaker \textit{et al.} \cite{Khandaker}} & {\scriptsize MISO, multiple Eves} & {\scriptsize Design artificial noise which is constructive to the legitimate receiver and destructive to the eavesdroppers.}\\ \hline
    \end{tabular}
\end{table}

\subsection{Perfect MCSI and Statistical ECSI}

While availability of perfect ECSI at the transmitter may be practical in some limited scenarios, in general, it is highly challenging to obtain it instantaneously. A more practical assumption regarding CSIT is the availability of perfect MCSI along with the statistical ECSI at the transmitter. On the other hand, the precoder design in this scenario is not as effective as those carried out with the perfect knowledge of both channels. In other words, with the absence of the instantaneous ECSI, precoding is not as forceful in suppressing the reception at the eavesdropper. For instance, transmission along the null-space of the eavesdropper's channel is not even possible in this scenario.

An example of secure transmission scheme under perfect MCSI and statistical ECSI is given in \cite{Nguyen} where the authors consider a MISOME channel and define a practical secrecy metric (instead of an information theoretic one), which quantifies the symbol error probability of the confidential data, and show that in the absence of artificial noise, secrecy diversity (i.e., the high-SNR slope of their defined metric) vanishes. This result underlines the importance of artificial noise injection in these scenarios.

Artificial noise-aided transmit precoding strategies with the objective of maximizing the secrecy rate are proposed in \cite{Bashar0}--\hspace*{-4px}\cite{TWC}. In \cite{Bashar0}, the authors employ naive beamforming along with the artificial noise injection while considering single-antenna receivers. The strategy proposed in \cite{Wu}, on the other hand, relies on an iterative maximization of an approximation to the instantaneous secrecy rate. In both of these studies, it is shown that the optimal schemes allocate only a fraction of the total power for signal transmission at high SNRs. Hence, the remaining power is employed for artificial noise injection.
It is shown in \cite{ISIT} that jointly optimizing the precoder matrix and portion of the power allocated to the artificial noise outperforms the solutions, which rely on optimizing the precoder only. Moreover, inspired by the idea proposed in \cite{GAN}, a generalized artificial noise aided transmission is introduced in \cite{TWC}, which guarantees high secrecy rates for the scenarios with $N_t < N_E$ for which injection of artificial noise along the null-space of the main channel is not possible. A summary of the existing solutions for the scenarios with perfect MCSI and statistical ECSI at the transmitter is given in Table \ref{tbl:precodingSECSI}.

\begin{table}[t!]
	\centering
	\caption{Precoder design with perfect MCSI and statistical ECSI.}
	\label{tbl:precodingSECSI}
	\begin{tabular}{|p{2.7cm}|p{1.7cm}|p{8.3cm}|}
		\hline
		{\scriptsize Authors} & {\scriptsize System Model} & {\scriptsize Contributions}                                                                                                                                   \\ \hline \hline
		{\scriptsize T. V. Nguyen \textit{et al.} \cite{Nguyen}}  & {\scriptsize MISOME} & {\scriptsize Define a weak notion of secrecy based on symbol error probability and quantify the impact of artificial noise injection with beamforming on diversity order. }                                                                                   \\ \hline
		{\scriptsize 	S. Bashar \textit{et al.} \cite{Bashar0}} & {\scriptsize MISOSE}& {\scriptsize  Employ a naive beamforming with the aid of MCSI only along with a power optimization using perfect MCSI and statistical ECSI. The excess power is used for artificial noise injection in the null-space of $\textbf{H}_B$.}\\ \hline
		{\scriptsize 	Y. Wu \textit{et al.} \cite{Wu}} & {\scriptsize MIMOME} & {\scriptsize Develops an iterative algorithm for precoder design and allocate the excess power to inject artificial noise along the null-space of $\textbf{H}_B$.}\\ \hline
		{\scriptsize S. Rezaei Aghdam and T. M. Duman \cite{TWC}} & {\scriptsize MIMOME} & {\scriptsize Propose a joint precoder and (generalized) artificial noise optimization algorithm.}\\ \hline
	\end{tabular}
\end{table}


\subsection{Perfect MCSI Only}

In the scenarios with a passive eavesdropper, a realistic assumption is that the transmitter does not know anything about the eavesdropper's channel. Under this assumption, \cite{Kalantari_conf} and \cite{Kalantari_jour} propose a secure transmission strategy referred to as directional modulation, in which the amplitude and the phase of the transmit signal are adjusted by varying the length of the reflector antennas for each symbol. This scrambles the PSK symbols in all the directions other than that of the legitimate receiver. Other strategies for securing communications without the knowledge of the eavesdropper's channel are proposed in \cite{THP} and \cite{Fakoorian}. Zhang \textit{et al.} develop a Tomlinson-Harashima precoding in \cite{THP} where the transmitter allocates part of its power in order to achieve a target mean squared error (MSE) at the legitimate receiver, and the remaining power is used to transmit artificial noise to degrade the eavesdropper’s reception. In \cite{Fakoorian}, a secure space-time block coding (STBC) scheme is proposed, which enables the legitimate receiver to perform separable decoding, while requiring an exhaustive maximum likelihood (ML) detection at the eavesdropper. Furthermore, the authors combine this scheme with artificial noise injection to ensure a high uncoded BER at Eve. In \cite{Hamamreh2016}, the authors propose a practical precoded orthogonal STBC method for MISO wireless networks where space-time codewords are precoded by matrix that minimizes the error rate at the legitimate receiver and a pre-equalization step is added to take security requirements into account. Table \ref{tbl:precodingMCSIonly} summarizes the existing transmit signal design approaches under perfect MCSI and no ECSI at the transmitter.

%


\begin{table}[h!]
	\centering
	\caption{Transmit signal design with perfect MCSI and no ECSI.}
	\label{tbl:precodingMCSIonly}
	\begin{tabular}{|p{2.7cm}|p{2cm}|p{8cm}|}
		\hline
		{\scriptsize Authors} & {\scriptsize System Model} & {\scriptsize Contributions}                                                                                                                                   \\ \hline \hline
		{\scriptsize A. Kalantari \textit{et al.} \cite{Kalantari_conf}, \cite{Kalantari_jour}} & {\scriptsize MIMOME}& {\scriptsize Steer the array beam such that the phase of the PSK modulated signal is scrambled in directions other than Bob.}\\ \hline
		{\scriptsize L. Zhang \textit{et al.} \cite{THP}} & {\scriptsize MIMOME} & {\scriptsize Propose a nonlinear Tomlinson-Harashima precoding along with along with artificial noise injection.}\\ \hline
        {\scriptsize  S. A. A. Fakoorian \textit{et al.} \cite{Fakoorian}} & {\scriptsize MIMOME $N_t = N_B = 2$, $N_E \geq 1$} &  {\scriptsize Design a STBC which results in a high BER for Eve and a low BER for Bob.}\\ \hline
        {\scriptsize  J. M. Hamamreh \textit{et al.} \cite{Hamamreh2016}} & {\scriptsize MISOSE $N_t=4, N_B=N_E=1$} &  {\scriptsize Propose a precoded orthogonal STBC which minimizes BER at Bob and is capable of satisfying security requirements.}\\ \hline
	\end{tabular}
\end{table}

\subsection{Statistical MCSI and ECSI}

While most of the existing physical layer security solutions rely on the assumption that the transmitter is capable of estimating at least the instantaneous main CSI, in some scenarios (e.g., for fast fading channels), it may be difficult for the transmitter to track the rapidly varying channel coefficients. For these cases, the impact of the discrete inputs on the achievable secrecy rates is analyzed in \cite{Girnyk_ISITA}. Under the assumption that both channels are doubly correlated, using the knowledge of transmit and receive correlation matrices, transmit signal design algorithms are proposed in \cite{PIMRC17} to maximize the resulting achievable secrecy rates. The results in \cite{PIMRC17} reveal that jointly optimizing the precoder matrix and artificial noise results in increased achievable secrecy rates with respect to precoding without artificial noise injection. The existing studies on MIMO wiretap channels with statistical CSI are summarized in Table \ref{tbl:precodingSCSIonly}.

\begin{table}[b!]
	\centering
	\caption{Transmit signal design with statistical MCSI and ECSI.}
	\label{tbl:precodingSCSIonly}
	\begin{tabular}{|p{2.7cm}|p{1.7cm}|p{8.3cm}|}
		\hline
		{\scriptsize Authors} & {\scriptsize System Model} & {\scriptsize Contributions}                                                                                                                                   \\ \hline \hline
		{\scriptsize M. Girnyk  \textit{et al.} \cite{Girnyk_ISITA}} & {\scriptsize MIMOME}& {\scriptsize Derive large system approximation for secrecy rate under arbitrary channel inputs and quantify it for Gaussian and finite-alphabet inputs.}\\ \hline
		{\scriptsize S. Rezaei Aghdam and T. M. Duman \cite{PIMRC17}} & {\scriptsize MIMOME} & {\scriptsize Propose precoding and artificial noise injection approaches with the aid of the knowledge on correlation matrices.}\\ \hline
	\end{tabular}
\end{table}

We now compare the secrecy rates achieved by different transmit signal design approaches under different CSIT assumptions. In particular, we overview the results for the scenarios with perfect or statistical CSI at the transmitter. When perfect CSI is available, each element of $\textbf{H}_B$ and $\textbf{H}_E$ are modeled as independent and identically distributed (i.i.d.) with $\mathcal{CN}(0,\sigma_B^2)$ and $\mathcal{CN}(0,\sigma_E^2)$, respectively.  When considering the statistical CSI corresponding to the main channel or the eavesdropper's channel, a commonly adopted model in the literature employs doubly correlated channels with
\begin{equation}\label{correlated}
\textbf{H}_B= \boldsymbol{\Psi}_{rB}^{1/2} \hat{\textbf{H}}_B \boldsymbol{\Psi}_{tB}^{1/2},~~~~~
\textbf{H}_E= \boldsymbol{\Psi}_{rE}^{1/2} \hat{\textbf{H}}_E \boldsymbol{\Psi}_{tE}^{1/2},
\end{equation}
where the elements of $\hat{\textbf{H}}_B$ and $\hat{\textbf{H}}_E$ are i.i.d., and the transmitter is capable of acquiring the transmit and receive correlation matrices, i.e., $\boldsymbol{\Psi}_{tb}$, $\boldsymbol{\Psi}_{te}$,  $\boldsymbol{\Psi}_{rb}$ and $\boldsymbol{\Psi}_{re}$, from the long-term statistics of the channel. The perfect instantaneous CSI is available at the receivers, including the legitimate receiver and the eavesdropper.

\begin{figure}[t!]
	\begin{center}
		\begin{tabular}{c}
			\mbox{\psfig{figure=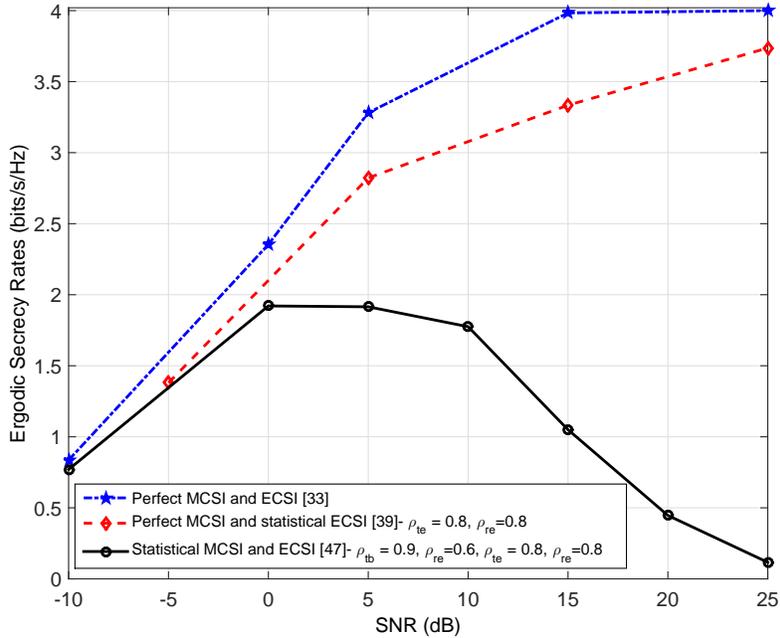,width=12cm}}
		\end{tabular}
	\end{center}
	\caption{Secrecy rates with different transmit signal design algorithms under different CSIT assumptions.}\label{diff_csit}
\end{figure}

Fig. \ref{diff_csit} demonstrates the achievable secrecy rates by different transmission schemes over a MIMOME wiretap channel with $N_t = 4$ and $N_B = N_E = 2$. Correlation matrices are assumed to have exponentially decaying entries as 
\begin{align}
[{\boldsymbol{\Psi}_{t}}(\rho)]_{ij} = {\rho_{t}}^{|i-j|},~~~~ &i,j = 1, 2, \ldots, N_t,\\
[{\boldsymbol{\Psi}_{r}}(\rho)]_{mk} = {\rho_{r}}^{|m-k|},~~~~ &m,k = 1, 2, \ldots, N_B ~(\text{or, } N_E).
\end{align}

It is demonstrated that the precoder design using a gradient descent optimization approach \cite{Wu}, \cite{TWC} is a promising strategy for maximizing the secrecy rates. Fig.  \ref{diff_csit} also reveals the importance of CSIT. When perfect CSI corresponding to the eavesdropper is available at the transmitter, it is possible to construct precoders, which are aligned with the null-space of the eavesdropper's channel. By this means, the achievable rate of the eavesdropper can be suppressed to zero. Furthermore, even when only the statistical CSI of the eavesdropper along with perfect CSI of the main user is available, high secrecy rates are still achievable with the aid of appropriately designed artificial noise injection strategies. However, in the absence of instantaneous CSI, the precoder matrices designed with the knowledge of correlation matrices are not capable of preventing the ergodic secrecy rates from dropping to zero at high SNRs. In other words, since statistical CSI does not provide sufficient degrees of freedom to take advantage of fading for improving secrecy, the behavior is similar to that in AWGN channels described in Section \ref{sec:p2pSISO}.

\subsection{Transmit Signal Design Based on Channel Reciprocity}

A group of transmission strategies in the literature of physical layer security rely on the assumption of channel reciprocity, that is, the transmitter and the legitimate receiver observe the same channel, simultaneously. For the systems working in time division duplex (TDD) mode, this property can be used to allow the transmitter to obtain the MCSI using the pilots transmitted from the legitimate receiver. In \cite{XLi}, a secure transmission scheme is proposed over a MIMO wiretap channel under the assumption that no training signals are transmitted by Alice, and hence, no CSI is available at Bob or Eve. Alice realizes an orthogonal space-time block coded PSK transmission and uses the CSI she obtains from the reverse channel estimation to design a phase shifting precoder, which compensates the phase shift by the channel. Therefore, Bob can recover the transmitted messages with no need to CSI. On the other hand, Eve who observes an independent channel can only recover the information with non-coherent detection, ensuring that this strategy provides positive secrecy rates over MIMO wiretap channels.

In the scenarios with channel reciprocity, when channel estimation is carried out both in the forward (from Alice to Bob) and the reverse (from Bob to Alice) directions, the channel between Alice and Bob can serve as a source of common randomness, which can further improve secrecy. This common randomness is used in \cite{Furqan} for generation of a random phase sequence, which is then used to manipulate the transmitted symbols (by simply multiplying them with the obtained random phase terms). Since the eavesdropper is unable to find out these random phase values, an enhanced secrecy is attained. In \cite{T_Allen_rec}, the authors propose a secure Alamouti scheme using channel reciprocity. In their proposed approach, after performing a two-way probing, the legitimate users exchange a secret key based on their received signal strength indicator, and this key is used for rotation of the PSK constellation from one codeword to the next. A cross-layer analysis of secure MIMO transmission is given in \cite{T_Dean}. It is shown that when using the main channel as a (not necessarily secret) key between the legitimate users, complexity of the eavesdropper's decoding of the legitimate receiver's message is as hard as solving standard lattice problems, and it grows exponentially as a factor of number of transmit antennas.

\subsection{Performance-Complexity Trade-off in Transmit Signal Design}
	
The iterative algorithms proposed in \cite{Wu} and \cite{TWC} for optimizing the precoder matrix are shown to offer promising results in terms of the achievable secrecy rates. However, their implementation complexity may be too high for some practical applications. There are two main reasons for this high computational complexity. First, both the mutual information and its gradient lack closed-form expressions. Therefore, Monte Carlo simulations need to be utilized to evaluate them, which require averaging over a large number of noise realizations in order to maintain a high level of accuracy. More importantly, evaluation of the mutual information and the minimum MSE (MMSE) term involves additions over the modulation signal space, which grows exponentially with the number of transmit antennas, resulting in a prohibitively high computational complexity when the number of transmit antennas is large.

Different strategies are proposed for lowering the computational complexity associated with the transmit signal design for secrecy under the finite-alphabet input assumption. The authors in \cite{Wu} and \cite{ISIT} derive closed-form approximations for the secrecy rate expressions using bounds on mutual information. While the former employs the bounds given in \cite{Zeng}, the latter approximates the mutual information using the cut-off rate expression. The authors in \cite{Girnyk_Spr}-\hspace{-1px}\cite{Girnyk15} derive asymptotic secrecy rates in a regime where the numbers
of antennas at the transmitter and both receivers grow infinitely large with a constant ratio. Maximization of this expression is shown to yield a satisfactory performance even in the cases of small number of antennas. The authors in \cite{Wu17} and \cite{TWC} propose per-group precoding strategies under perfect and partial CSIT assumptions, respectively. In these works, the channels are diagonalized, and the precoder matrices are designed by grouping the transmit antennas and designing the precoder matrix for each group, independently.

\subsection{Secure Spatial Modulation}

Spatial modulation and space shift keying (SSK) are relatively new MIMO transmission schemes, which rely on encoding information via active antenna indices. Similar to the amplitude or phase modulation schemes, in spatial modulation and SSK, the channel inputs are drawn from discrete and finite sets. More specifically, in spatial modulation \cite{Mesleh}, the data bits are mapped onto two information carrying blocks: 1) an antenna index selected from the set of transmit antennas, and 2) a symbol drawn from a standard constellation. Space shift keying is a special case of spatial modulation where no amplitude of phase modulation is employed, i.e., the transmit antenna index serves as the only information carrying unit \cite{Jeganathan}.

Different studies explore the advantages offered by spatial modulation over state-of-the-art MIMO schemes (see \cite{Prc13} for a comprehensive overview). Along with different application areas of spatial modulation and SSK, studying these transmission schemes in the context of physical layer security has been of recent interest. Secrecy capacity of spatial modulation for the scenarios with two transmit antennas is studied in \cite{Sinanovic}, and the results are extended to MISOSE and MIMOME scenarios in \cite{Guan} and \cite{Conf}, respectively. 

\begin{table}[]
	\centering
	\caption{Transmit signal design for secure spatial modulation and SSK.}
	\label{tbl:SMandSSK}
	\begin{tabular}{|p{2.7cm}|p{3cm}|p{7cm}|}
		\hline
		Paper         & Model                & Contributions                                                                                                                                   \\ \hline \hline
		{\scriptsize S. Sinanovic \textit{et al.} \cite{Sinanovic}} & {\scriptsize $N_t=2$, $N_B=N_E=1$, No CSI at the transmitter} & {\scriptsize Characterize secrecy rates for spatial modulation and SSK transmission.} \\ \hline
		{\scriptsize X. Guan \textit{et al.} \cite{Guan}} & {\scriptsize MISOSE} &  {\scriptsize Evaluate the secrecy rates by deriving mutual information terms and propose a precoding with the aid of full CSI.} \\ \hline
		{\scriptsize S. Rezaei Aghdam and T. M. Duman \cite{Conf}} & \scriptsize{MIMOME}, No CSI at the transmitter & {\scriptsize Derive expressions for achievable secrecy rates with spatial modulation and SSK schemes.} \\ \hline
		{\scriptsize S. Rezaei Aghdam and T. M. Duman \cite{Lett}} & \scriptsize{MIMOME}, Full CSI at the transmitter & {\scriptsize Develop an iterative precoder design algorithm.} \\ \hline
		{\scriptsize F. Wu \textit{et al.} \cite{SPSM}} & {\scriptsize MIMOME, $N_t>N_B$, MCSI at the transmitter} &  {\scriptsize Propose a precoder and artificial noise design.} \\ \hline
	    {\scriptsize L. Wang \textit{et al.} \cite{Wang_SM}} & {\scriptsize MIMOME, $N_t>N_B$, MCSI at the transmitter} &  {\scriptsize Characterize the secrecy rates of artificial noise aided spatial modulation.} \\ \hline
	    {\scriptsize Z. Huang \textit{et al.} \cite{QSM}} & {\scriptsize MIMOME,  MCSI at the transmitter.} & {\scriptsize Propose an artificial noise aided secure quadrature spatial modulation which requires low-complexity hardware.}\\ \hline
	\end{tabular}
\end{table}

Different secure spatial modulation based transmission schemes are also proposed with the aid of MCSI and ECSI at the transmitter, and for systems with channel reciprocity. A precoder optimization algorithm for maximizing secrecy rates with SSK transmission using full CSI is proposed in \cite{Lett}. The authors in \cite{SPSM} provide a precoder design and artificial noise injection approach for securing spatial modulation for the scenarios with passive eavesdroppers (with no ECSI at the transmitter). The secrecy rates of spatial modulation with artificial noise injection are characterized in \cite{Wang_SM}. While the conventional artificial noise injection (similar to what is considered in \cite{Wang_SM}) requires multiple antennas to be activated, a novel secure artificial noise aided transmission is introduced in \cite{QSM}, which requires only two antennas to be activated at each time instance. Existing secure transmit signal design approaches for spatial modulation and SSK are summarized in Table \ref{tbl:SMandSSK}.

For the scenarios where channel reciprocity holds, a transmit preprocessing technique is proposed in \cite{Li_SM} under the assumption that the training sequences are transmitted from Bob to Alice, and no CSI is available at the receiver nodes. The authors in \cite{Globecom} propose a dynamic antenna index assignment based on the main channel when training is carried out at both directions, which prevents an eavesdropper from obtaining the antenna index assignment pattern. 

\section{Multi Carrier Wiretap Channel with Discrete Inputs}
\label{sec:multicarrier}
\subsection{OFDM Wiretap Channel}
\begin{figure}[b!]
	\begin{center}
		\begin{tabular}{c}
			\mbox{\psfig{figure=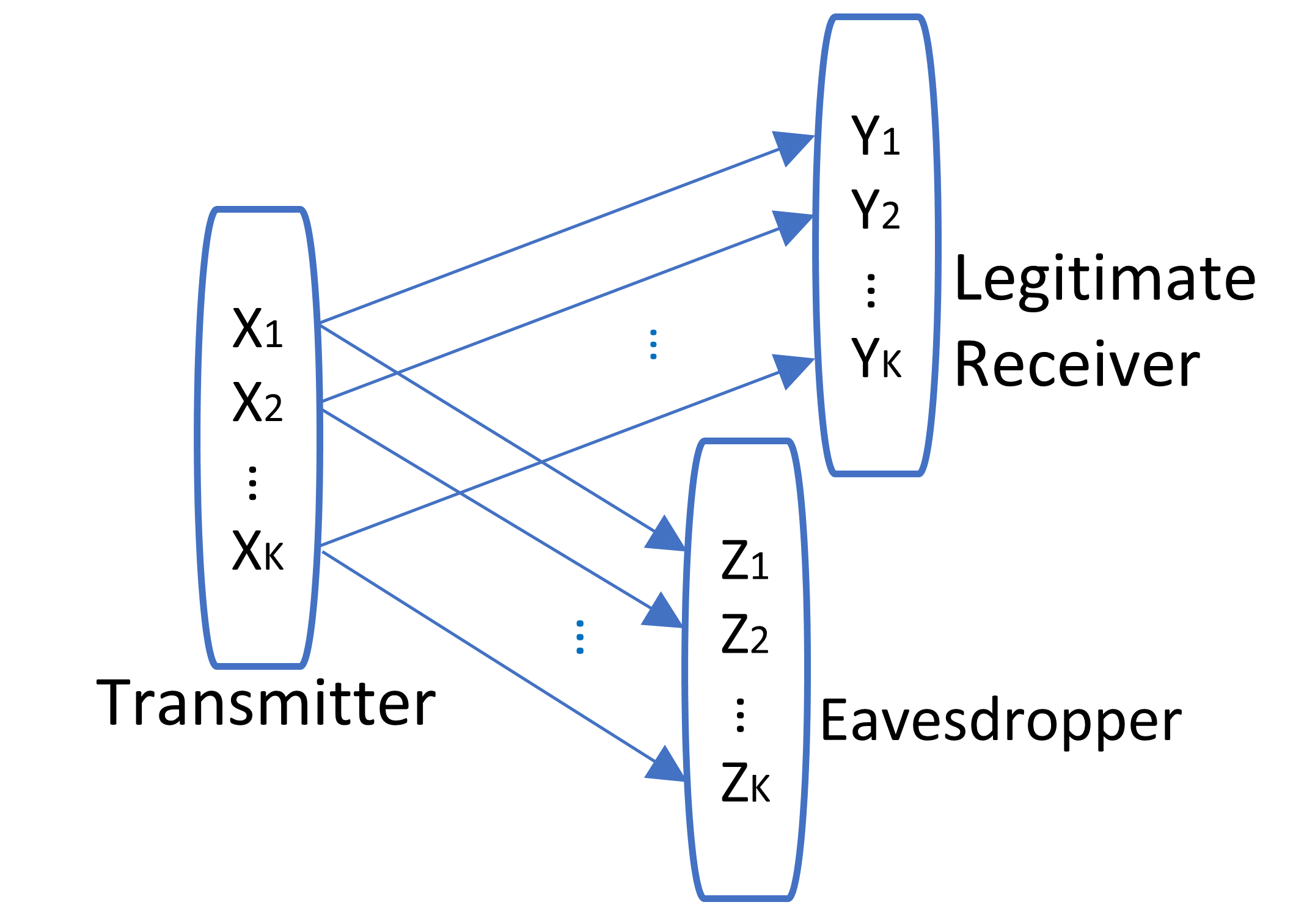,width=8cm}}
		\end{tabular}
	\end{center}
	\caption{OFDM wiretap channel modeled as set of independent parallel channels.}\label{para}
\end{figure}

Orthogonal frequency-division multiplexing (OFDM) provides robustness against multipath channel fading and offers flexibility in resource allocation. These features have made OFDM a widely used technique in recent wireless standards including 4G and 5G wireless systems. In the literature of physical layer security, OFDM can be modeled as a set of independent parallel channels.
It is proved in \cite{parallel} that the secrecy capacity of $K$ independent parallel channels (as depicted in Fig. \ref{para}) is achieved when each channel achieves its individual secrecy capacity. For parallel Gaussian wiretap channels, the optimal input is Gaussian distributed \cite{Renna_cap}. However, a more practical study of OFDM wiretap channel is possible via assuming that the data symbols belong to finite constellations.

An initial study on parallel Gaussian wiretap channels under finite-alphabet inputs is reported in \cite{Rodrigues} where the authors characterize the secrecy capacity with PAM inputs. They also provide a Mercury-waterfilling interpretation of the optimal power allocation strategy. It is observed in \cite{Rodrigues} that dissimilar to standard parallel Gaussian channels (with no constraints on the input distribution), it may not be optimal to use all available power for PAM inputs. In \cite{Renna}, the OFDM wiretap channel is studied with QAM inputs. After quantifying the loss in the secrecy rates with respect to the scenarios with Gaussian inputs, a bit loading strategy is proposed to minimize this loss by assigning an appropriate number of bits to each subchannel.

The sensitivity of OFDM based transmissions to frequency synchronization errors can be exploited to provide secure transmission as well. The authors in \cite{Yusuf} propose a secure OFDM transmission over reciprocal channels, which relies on
inducing carrier frequency offset that is pre-compensated (using the transmitter's knowledge on the MCSI) in such way that it is received without inter-carrier interference at the legitimate receiver, while the reception at the eavesdropper receiving the signal through an independent channel is considerably degraded. In \cite{HaoLi}, an eavesdropping resilient OFDM scheme is developed over reciprocal channels where Alice performs a subcarrier interleaving using her knowledge on the instantaneous MCSI. The legitimate receiver who is also capable of acquiring the MCSI can obtain the interleaving pattern and de-interleave the received signals. In contrast, the eavesdropper cannot derive the interleaving pattern initiated by the transmitter, and as a result, fails to acquire the transmitted information.

Cyclic prefix is used in OFDM systems to avoid inter carrier interference (ICI) incurred by multi-path channels. However, using cyclic prefix makes the transmission vulnerable to cyclostationarity based attacks \cite{cyclostationary}. This temporal degrees of freedom offered by the cyclic prefix can be used for artificial noise injection so as to improve secrecy over OFDM wiretap channels. In \cite{HQin}, a single-antenna OFDM wiretap channel is considered and  an artificial noise signal is inserted into the time-domain signal (over the cyclic prefix samples). A framework for joint optimization of the power allocated to each subcarrier and artificial noise covariance matrix is also proposed. In \cite{Akitaya}, Akitaya \textit{et al.} study the application of time-domain artificial noise injection in MIMOME OFDM systems. The results in \cite{HQin} and \cite{Akitaya} demonstrate that this approach is efficient for improving the secrecy rates in OFDM wiretap channels. 

Efficient resource allocation for secure OFDM systems is formulated as a convex optimization problem in \cite{Kwan2012}, and solved assuming that the transmitter uses artificial noise to combat an eavesdropper equipped with multiple antennas. Li et al. \cite{Li2013} propose scrambling of data in the time domain after the IFFT operation in order to improve physical layer security for OFDM signals. This, in fact, corresponds to sending modified constellation points (in terms of phase and amplitude) on each sub-carrier. The authors in \cite{Hamamreh2017} introduce channel frequency response based pre-coder and post-coder designs for enhancing security in OFDM systems. Their solution relies on extracting orthonormal matrices via applying singular value decomposition (SVD) on the diagonal matrix of the legitimate user’s channel frequency amplitude and using it for shuffling OFDM sub-channels as well as for designing the frequency-based precoder
and post-coder. In \cite{Hajomer2018}, physical layer security is studied for passive OFDM-based optical networks where a scrambled precoding technique is utilized to improve the security. The authors in \cite{Yusuff} propose an efficient approach to enhance secrecy in OFDM systems that relies on employing an adaptive interleaver maximizing the overall diversity gain delivered to the legitimate receiver. Hamamreh et al. \cite{HAMAMREH2017B} propose an efficient physical layer security technique for transmission of OFDM-based waveforms by developing an optimal joint subcarrier index selection and adaptive interleaver design to enhance the security and reliability of 5G services. In order to provide QoS based security for OFDM systems, channel shortening (CS) equalizer coefficients are designed in \cite{Furkan2017} based on the legitimate receiver's channel. Specifically, CS is used at the transmitter in such a way that the effective channel ensures no ISI at Bob, while causing ISI and performance degradation at Eve.

Table \ref{tbl:OFDMbased} summarizes important results on the study of OFDM wiretap channels under the finite-alphabet input assumption. 

\begin{table}[]
	\centering
	\caption{OFDM wiretap channel with discrete inputs.}
	\label{tbl:OFDMbased}
	\begin{tabular}{|p{2.7cm}|p{2.7cm}|p{7.3cm}|}
		\hline
		Paper         & Model                & Contributions                                                                                                                                   \\ \hline \hline
		{\scriptsize M. R. D. Rodrigues \textit{et al.} \cite{Rodrigues}} & {\scriptsize Parallel Gaussian Wiretap Channels} & {\scriptsize Characterize secrecy rates with PAM inputs and derive optimal power allocation strategy.} \\ \hline
		{\scriptsize F. Renna \textit{et al.} \cite{Renna}} & {\scriptsize SISOSE} &  {\scriptsize Evaluate the secrecy rates for OFDM systems with QAM input constellations and provide a bit-loading strategy.} \\ \hline
		{\scriptsize M. Yusuf \textit{et al.} \cite{Yusuf}} & \scriptsize{SISOSE with reciprocal main channel} & {\scriptsize Suggest inducing carrier frequency offset which is pre-compensated in the legitimate receiver's direction.} \\ \hline
		{\scriptsize H. Li \textit{et al.} \cite{HaoLi}} & {\scriptsize SISOSE with reciprocal main channel}  & {\scriptsize Propose subcarrier interleaving based on the MCSI which prevents eavesdropper from de-interleaving the received signals.} \\ \hline
		{\scriptsize H. Qin \textit{et al.} \cite{HQin}} & \scriptsize{SISOSE} & {\scriptsize Introduce a time-domain artificial noise injection approach and propose a joint optimization of data power and artificial noise covariance matrix.} \\ \hline
		{\scriptsize T. Akitaya \textit{et al.} \cite{Akitaya}} & {\scriptsize MIMOME}  & {\scriptsize Study application time-domain artificial noise injection approach in multi-antenna scenarios.} \\ \hline
		{\scriptsize M. Yusuf \textit{et al.} \cite{Yusuff}} & \scriptsize{SISOSE with reciprocal main channel} & {\scriptsize Employ an interleaving whose pattern is adapted to the main channel.} \\ \hline
		{\scriptsize J. M. Hamamreh \textit{et al.} \cite{HAMAMREH2017B}} & {\scriptsize SISOSE with reciprocal main channel}  & {\scriptsize Develop a joint optimal subcarrier index selection and adaptive interleaving design to enhance the security and reliability} \\ \hline
	\end{tabular}
\end{table}

%
\subsection{Filter-Bank Multicarrier Wiretap Channel}

OFDM systems face important challenges when they are adopted for more complex networks. For instance, it is very difficult to establish full synchronization for the signals corresponding to different users in the base station in an orthogonal frequency division multiple access (OFDMA) network. Furthermore, OFDM introduces significant out of band interference to other users when it is used to transmit over a set of non-contiguous frequency bands as in cognitive radios where transmission is limited to certain portions of the band and 5G wireless systems \cite{Farhang2016}. 

Filter bank multicarrier (FBMC) modulation is shown to be able to resolve the above problems utilizing high quality filters for transmission, which makes it a potential candidate for future networks including 5G. With this motivation, recently, there has been some interest in the physical layer security aspects of FBMC based networks, as well. In \cite{Schellmann2014}, the authors present a filter hopping FBMC approach for security. They argue that the choices of filters for each symbol provide a major degree of freedom for the FBMC waveforms, which can be exploited for security purposes. Furthermore, a mismatch in the receive filter results in violation of the orthogonality of different subcarriers, and it causes the energy of a transmit symbol in a certain subcarrier to spread into the adjacent subcarriers. This effect is shown to be significant even for a slight mismatch of the filters between the transmitter and the receiver \cite{Schellmann2014}. Such sensitivity can be exploited for providing physical layer security. For instance, in a practical system, Alice and Bob can exchange their filter designs by exchanging keys or using a pre-shared sequence of filters. Then, the transmitter can change the filters continuously, and make it extremely difficult for an eavesdropper to find the exact filter being used, hence securing the transmission.

\section{Spread spectrum techniques for physical layer security}
\label{sec:SS}
One of the potential methods for improving physical layer security is spread spectrum (SS) communications. In this method, a signal is spread over a frequency band such that the resulting bandwidth is much wider than that of the original signal. Historically, this method has been first used in military applications to achieve transmission schemes with low probability of intercept (LPI). Two main approaches in SS are direct sequence spread-spectrum (DSSS) and frequency-hopping spread-spectrum (FHSS). DSSS is used to spread the transmitted data over multiple frequencies. Similar to a fast hopping system, FHSS constantly changes the carrier frequency several times per bit duration to correspond with a randomly selected channel \cite{SSBook}. In this way, it is extremely difficult for an eavesdropper to monitor the spread spectrum signals. One of the main advantages of the SS in comparison to traditional cryptography methods (where the key space is very large) is that the key space is limited by the range of carrier frequencies.

Code-division multiple access (CDMA), which is based on spread spectrum communications has also been studied for secure communications in the literature. One important type of CDMA systems is direct-sequence CDMA (DS-CDMA) where users utilize different spreading codes to distinguish their signals while they share the same channel \cite{CDMABook}. For a secure DS-CDMA system, the signal from the transmitter is spread using a code sequence. In order to avoid an eavesdropper to detect the signal, the spread signal is then scrambled using a pseudo-noise (PN) sequence. 

Physical layer security of CDMA systems is studied in \cite{Li2005} based on the use of relatively long PN scrambling sequences. In most cases, Walsh codes are used to generate channelization codes for this purpose as they are easy to generate. The authors in 
\cite{Li2005} propose a method to generate the scrambling sequences by using advanced encryption standard (AES) operation. Specifically, they design an AES-CDMA scheme secured against exhaustive key-search attacks utilizing AES with three different key sizes up to $256$.

\section{Constellation Constrained Multiuser and Cooperative Wiretap Channels}
\label{sec:multiuser}

The concept of physical layer security has also been extended to multi-user and cooperative networks in recent years. In addition to the numerous studies on the limits of secure communications for scenarios with more than two receivers, multiple transmitters and presence of  relays/jammers/helpers (see \cite{Mukherjee} for a survey), some recent focus has also been placed on the practical constellation-constrained secure transmission in these settings.
%

\subsection{Multiuser and Multi-Eve Networks}

\subsubsection{Broadcast Channel with Confidential Messages}
The achievable rate region for Gaussian broadcast channel with confidential messages is characterized for discrete inputs in \cite{Mheich}. In this scenario, the transmitter's objective is to send a common message to two receivers as well as to deliver a confidential message to one of them. Dissimilar to  \cite{Raghava} and \cite{Rodrigues} where standard constellations are considered, in \cite{Mheich}, the symbols are allowed to be arbitrary, and the achievable secrecy rate region is enlarged by optimizing the symbol positions and the probability distribution of the symbols. The shrinkage of the rate region due to employing PAM inputs is quantified with respect to the optimal Gaussian inputs, and a number of important behavioral differences between these cases are highlighted.

\begin{figure}[b!]
	\begin{center}
		\begin{tabular}{c}
			\mbox{\psfig{figure=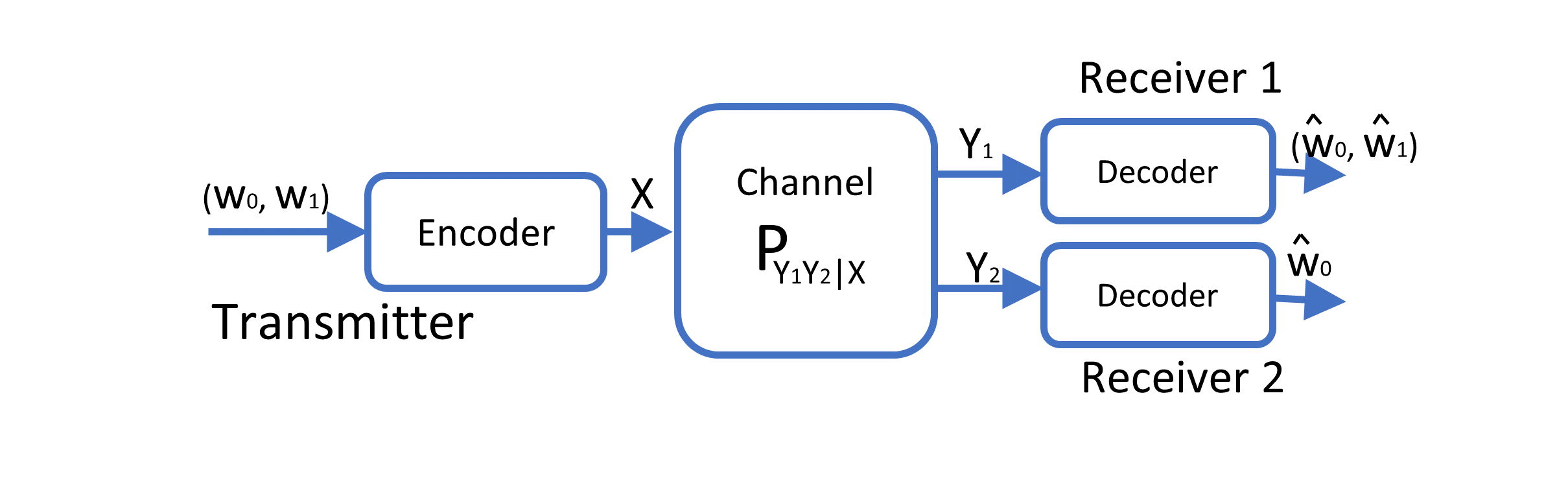,width=8cm}}
		\end{tabular}
	\end{center}
\vspace{-2em}
	\caption{Broadcast channel with confidential messages \cite{Mheich}.}\label{mheic}
\end{figure}

%
%
\subsubsection{Multiple Access Channels}
\begin{figure}[t!]
	\begin{center}
		\begin{tabular}{c}
			\mbox{\psfig{figure=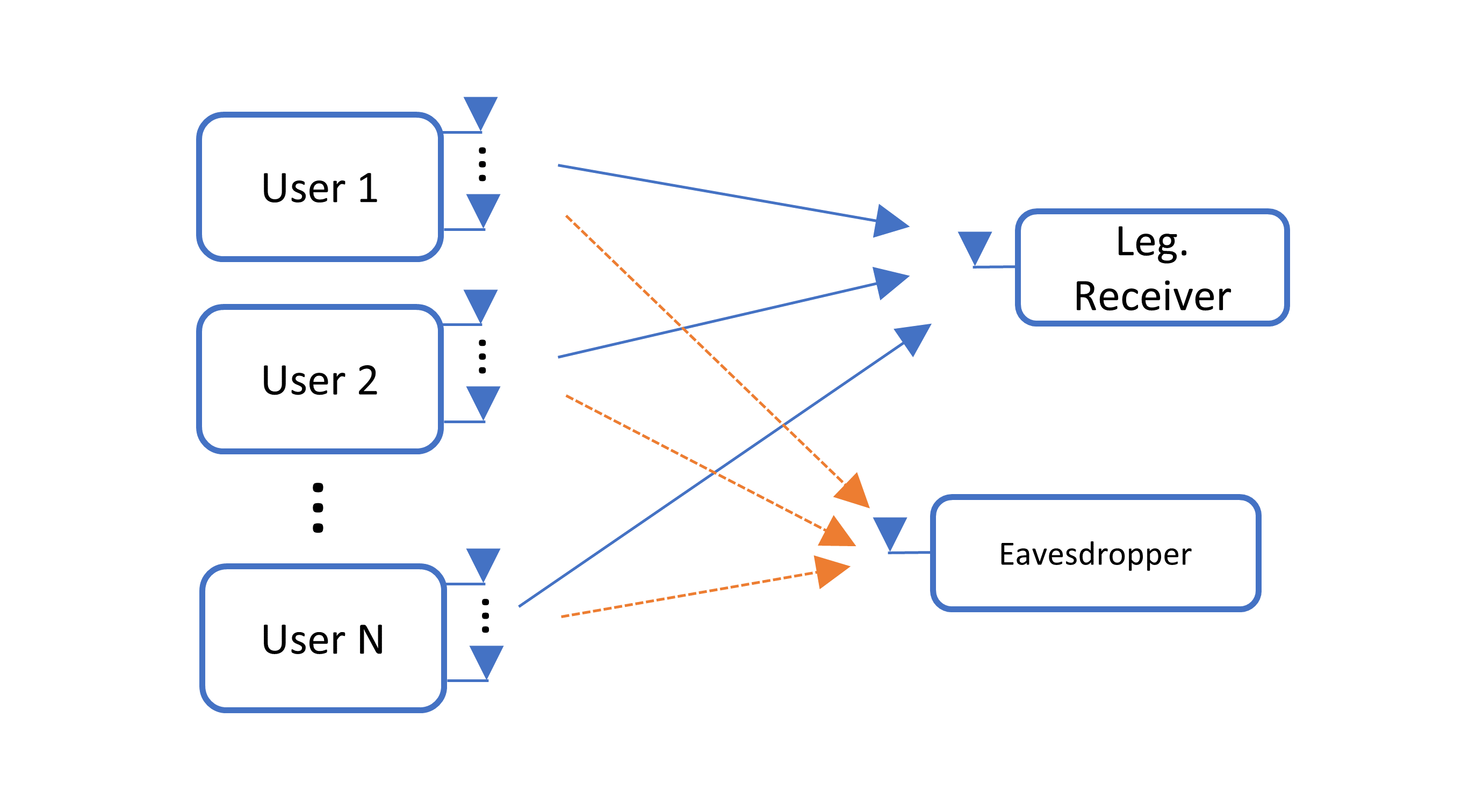,width=8cm}}
		\end{tabular}
	\end{center}
\vspace{-2em}
	\caption{Multiple access wiretap channel studied in \cite{MAC_Allen}.}\label{MAC}
\end{figure}

The authors in \cite{MAC_Allen} focus their attention on the multiple access wiretap channels (as depicted in Fig. \ref{MAC}) under the assumption that the users employ the Alamouti STBC scheme. Specifically, they consider a case where multiple transmitters communicate with one legitimate receiver in the presence of an eavesdropper. Under the assumption that the transmitter has the knowledge of the legitimate channel and does not know the eavesdropper's channel gain, an artificial noise injection is proposed along the null-space of the main channel, which is shown to give rise to considerable improvements in the secrecy sum-rate.

\subsubsection{Cognitive Radio Networks}

\begin{figure}[t!]
	\begin{center}
		\begin{tabular}{c}
			\mbox{\psfig{figure=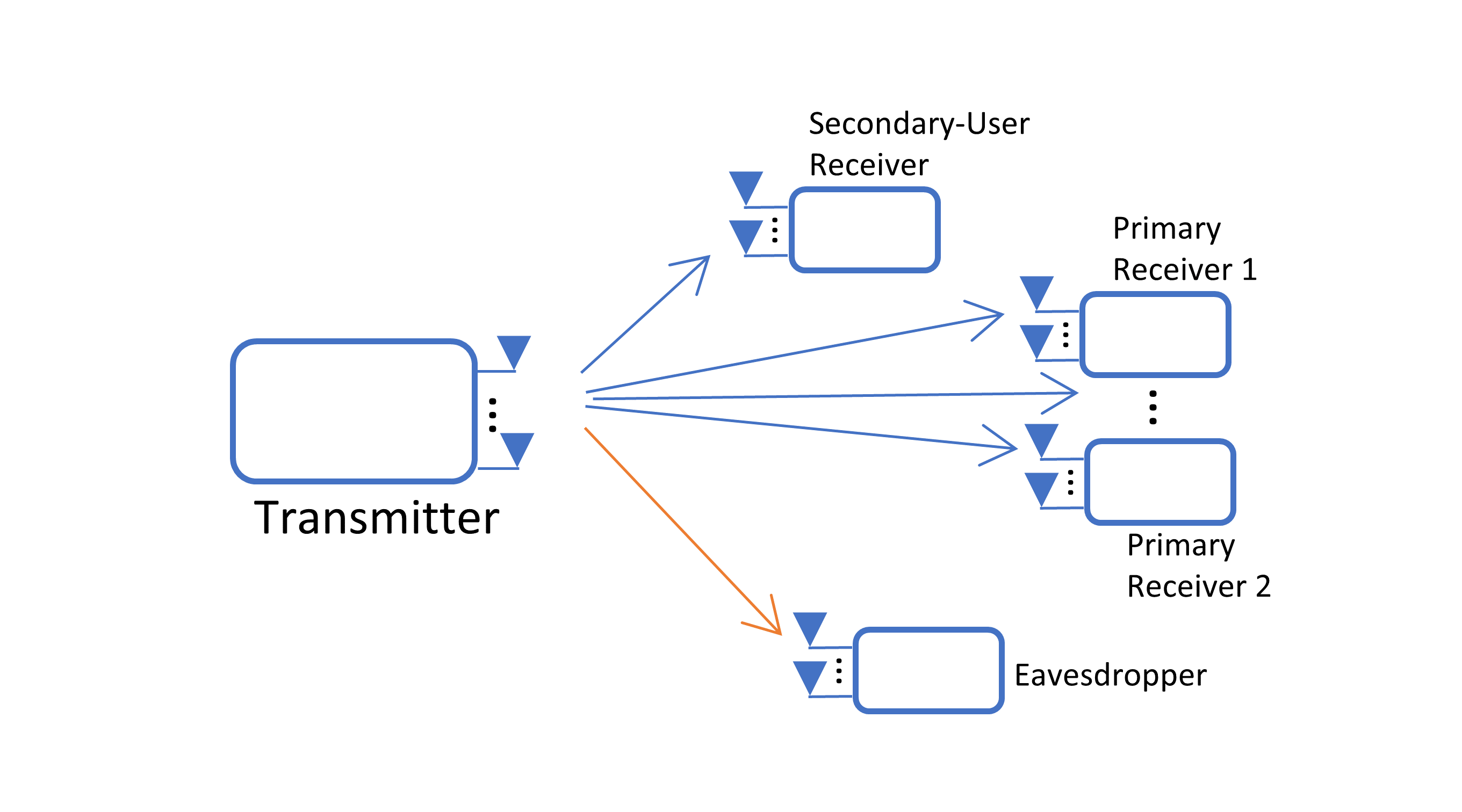,width=8cm}}
		\end{tabular}
	\end{center}
	\vspace{-2em}
	\caption{Cognitive multiple-access wiretap channel studied in \cite{Zeng16}.}\label{cogn}
\end{figure}

The linear precoder design for the cognitive multiple-access wiretap and the cognitive multi-antenna wiretap channels with finite-alphabet inputs are studied in \cite{Jin16} and \cite{Zeng16}, respectively. In \cite{Jin16}, the authors consider a setup in which two secondary-user transmitters  
are communicating with a secondary-user receiver in presence of an eavesdropper and under interference threshold constraints at the primary users. A two-layer precoding algorithm is proposed using statistical CSI at the transmitters to maximize the ergodic secrecy sum rate. In \cite{Zeng16}, on the other hand, the authors study a scenario where a multi-antenna secondary-user transmitter communicates with a multi-antenna secondary-user receiver, and the communication is wiretapped by a multi-antenna eavesdropper (Fig. \ref{cogn}). In this work, the precoder matrix is optimized using an iterative algorithm (assuming statistical CSI of all the channels) so that the secrecy rate of the secondary user is maximized while power leakage to the primary user sharing the same frequency spectrum is controlled.

\subsubsection{Multi-Eavesdropper Scenarios}

The authors in \cite{Cao18} study a setup comprised of a source, a legitimate receiver and multiple multi-antenna
eavesdroppers. They first obtain an expression for an achievable secrecy rate, and then, under the full CSI assumption, they optimize the transmit power and the beamforming vector. The problem of power allocation for secrecy over MIMO wiretap channels with multiple eavesdroppers is studied in \cite{Vishwakarma14}. Under the assumption that the transmitter has perfect MCSI and statistical ECSI, the proposed power allocation strategy gives non-zero secrecy rates at high transmit powers. This is important because in the absence of this power control strategy, secrecy rate decreases with increasing transmit power, and it drops to zero when the eavesdroppers' SNRs get higher than a certain value.
%

\subsection{Relay Channels and Cooperative Communications}

Cooperation serves as an efficient method for enhancing secrecy in wireless networks. In this context, along with the studies conducted on secrecy capacity in various cooperative scenarios, a number of recent studies focus on finite-alphabet inputs as the signaling scheme. In this context, the secrecy rates achieved with decode-and-forward (DF) relay beamforming under the finite-alphabet input assumption are characterized in \cite{Vishwakarma}. It is shown that by optimizing the source power and also the relay weights using the global CSI, it is possible to prevent the secrecy rates from dropping to zero. The authors in \cite{Cao17} consider a cooperative jamming network in presence of multiple eavesdroppers, and develop a secure transmission scheme with finite-alphabet inputs, which relies on a joint optimization of the artificial noise (injected in the null-space of relay-destination links) and power allocation among the source and the relays. The study in \cite{Cao_Helper} considers secure transmission over a MIMOME setup with the aid of a multi-antenna helper, which transmits a jamming signal to degrade the signal received by the eavesdropper. Similar to \cite{Cao17}, a framework is provided for joint optimization of the precoder matrix and power allocation between the transmitter and the helper. Furthermore, low-complexity transmit signal design schemes are developed for both low and high SNR regimes. The authors in \cite{Furqan_SW} focus on the problem of secure transmission with cooperative jamming via an untrusted relay. They propose a power efficient secure DF-based cooperative communication technique and demonstrate the efficacy of their scheme via simulations using practical constellation-constrained inputs.

\begin{table}[]
	\centering
	\caption{Multi-user, multi-eavesdropper and cooperative secure transmission with discrete inputs.}
	\label{tbl:multi}
	\begin{tabular}{|p{2.5cm}|p{3cm}|p{7.6cm}|}
		\hline
		Paper         & Network   & Contributions                                                                                                                                   \\ \hline \hline
		{\scriptsize Z. Mheich \textit{et al.} \cite{Mheich}} & {\scriptsize Broadcast channel with
			confidential message} & {\scriptsize Characterize secrecy rate region with PAM inputs and enhance it by optimizing symbol positions and the joint probability distribution.} \\ \hline
		{\scriptsize T. Allen \textit{et al.} \cite{MAC_Allen}} & {\scriptsize MAC Wiretap} &  {\scriptsize Develop a secure Alamouti transmission with artificial noise injection with perfect MCSI and statistical ECSI.} \\ \hline
		{\scriptsize J. Jin \textit{et al.} \cite{Jin16}} & \scriptsize{Cognitive MAC Wiretap} & {\scriptsize A two-layer precoding is proposed using statistical CSI to maximize secrecy sum-rate.} \\ \hline
		{\scriptsize W. Zeng \textit{et al.} \cite{Zeng16}} & {\scriptsize Cognitive Multi-Antenna Wiretap}  & {\scriptsize Propose an iterative precoder optimization algorithm with statistical CSI.} \\ \hline
		{\scriptsize K. Cao \textit{et al.} \cite{Cao18}} & {\scriptsize MISO Wiretap with Multiple Multi-Antenna Eavesdroppers}  & {\scriptsize Study joint optimization of beamforming vector and transmitting power with full CSI.} \\ \hline
		{\scriptsize S. Vishwakarma \textit{et al.} \cite{Vishwakarma14}} & {\scriptsize MIMO Wiretap with Multiple Eavesdroppers}  & {\scriptsize Employ a power allocation strategy with perfect MCSI and statistical ECSI.} \\ \hline
		{\scriptsize S. Vishwakarma \textit{et al.} \cite{Vishwakarma}} & {\scriptsize DF Relaying in Presence of Multiple Eavesdroppers}  & {\scriptsize Optimize the source power and the relay weights with the aid of full CSI such that the secrecy rate is maximized.} \\ \hline
		{\scriptsize K. Cao \textit{et al.} \cite{Cao17}} & {\scriptsize Cooperative Jamming Network in Presence of Multiple Eavesdroppers}  & {\scriptsize Provide a joint optimization of the artificial noise and the power allocated between the source and the relays.} \\ \hline	
	   {\scriptsize K. Cao \textit{et al.} \cite{Cao_Helper}} & {\scriptsize MIMOME and a Multi-Antenna Helper}  & {\scriptsize Propose a joint optimization of precoder and power allocation between the source and the helper} \\ \hline
	   {\scriptsize H. M. Furqan \textit{et al.} \cite{Furqan_SW}} & {\scriptsize Cooperative communications with untrusted relay}  & {\scriptsize Propose a destination-assisted jamming for improving secrecy and power efficiency.} \\ \hline

	\end{tabular}
\end{table}

Table \ref{tbl:multi} summarizes the recent research results of secure transmission over finite-input multi-user, multi-eavesdropper and cooperative channels.
%
%

\section{Physical Layer Security for Next Generation Wireless Systems}
\label{sec:5G}

With the advent of the next generation wireless communication systems such as 5G, a tremendous growth is expected in the number of connected devices. To meet the high traffic demand, various technologies have emerged during the past years. For instance, massive MIMO \cite{Larsson}, millimeter wave (mmWave) communication systems \cite{Rappaport}, full duplex transmission \cite{ZZhang} and non-orthogonal multiple access (NOMA) \cite{LDai} have attracted significant interest from researchers both in industry and academia.


Exploiting large antenna arrays and taking advantage of a spectrum from 30 GHz to 300 GHz away from the almost fully occupied spectral band, have proven to provide various benefits, which make these technologies key enablers for the next generation wireless systems and beyond. Along with the different aspects of massive MIMO and mmWave communications, investigating their potential in providing physical layer security has been of recent interest (see e.g., \cite{JWang}-\hspace{-0.1px}\cite{YZhu}). The problem of linear precoder design for large-scale (massive) MIMO wiretap channel under the finite-alphabet input assumption is investigated in \cite{Wu17} where GSVD-based precoder design is shown to achieve the maximal secrecy rate at high SNR values. The potential of large antenna arrays to realize secure directional modulation at mmWave frequencies is demonstrated in \cite{Valli13} where the proposed solution relies on driving only a subset of antennas in the array and choosing the switching configuration such that the desired modulation symbols are received at the legitimate receiver while scrambling the constellation in other (undesired) directions. The authors in \cite{JFan17} propose an antenna selection-aided secure mmWave transmission using a switched phased array architecture in large-scale transmit antenna arrays. We note that, different from the solution proposed in \cite{Valli13}, which requires modulation at the RF end, the scheme in \cite{JFan17} can use the traditional baseband modulation.

Offering several advantages such as high throughput, low latency and improved connectivity, NOMA is one of the key enabling technologies for next generation wireless networks. It relies on removing the orthogonality (which exists in the conventional multiple access techniques), and assigning a specific resource block (e.g., time or frequency) to more than one user. Investigating the security issues of NOMA has been a topic of recent interest. For instance, the secrecy performance of large-scale NOMA networks is studied in \cite{YLiu17} where their secrecy outage probability is derived for both single-antenna and multi-antenna transmitter scenarios. Proposing eavesdropping resilient transmission schemes with NOMA (especially, under a discrete signaling assumption) is an interesting research direction, which is left for future work as also stated in \cite{DXu}. 

Full duplex wireless communication is another promising technology, which is expected to lead to a considerable increase in spectral efficiency. Transmission via full duplex communications allows for simultaneous transmitting and receiving of information signal at the same time and within the same frequency band. In theory, the spectral efficiency can be doubled via full duplex transmission with respect to the conventional half duplex systems. The research on physical layer security in the scenarios where different nodes are capable of full duplex operation is a hot research topic (for an overview, see \cite[Section VII]{Wu_phy_sec}). However, only a limited number of the existing studies focus their attention to practical constellation-constrained scenarios. One example of such studies is reported in \cite{QLi16} where secure communications in presence of full duplex relays is investigated under the assumption that the transmitter employs an Alamouti-based rank-two beamforming scheme.

Waveform design for 5G has also been a topic of interest during the past years. In addition to an increased bandwidth efficiency and reduced inter-user interference, new waveforms may also provide gains in terms of secrecy. For instance, the authors in \cite{HamamrehOTDM} propose a secure waveform design for 5G wireless systems referred to as orthogonal transform division multiplexing (OTDM), which relies on replacing Fourier transform blocks in conventional OFDM systems by new blocks allowing for diagonalization of the main channel while degrading the eavesdropper's reception.

We conclude this section by noting that the relatively recent paradigm of Internet of things (IoT) is expected to offer connectivity to billions of devices and it will be a centerpiece in the next generation wireless technologies including 5G. In view of the fact that, IoT systems are expected to consist of a large number of devices, the conventional cryptographic solutions, which require key distribution and management, may be too complex to be implemented. Accordingly, physical layer security techniques (especially, the ones which possess a lower complexity) can play an important role in safeguarding IoT. A recent survey paper \cite{IoT} provides a thorough discussion of the potentials of physical layer security techniques and pinpoints a number of candidate schemes that can serve as an alternative or complement to the key-based solutions in securing IoT communications.



%
%

\section{Practical Wiretap Code Constructions}
\label{sec:Coding}

The achievability of the secrecy rates described in the previous sections is proved with the aid of random coding arguments, i.e., from an information theoretic point of view. However, in practice, one needs to realize secure communications with the aid of practical codes. In this regard, explicit and implementable code design for secrecy has also been a topic of recent interest in the literature. In particular, error-correcting codes have found a new role since the emergence of physical layer security. Aside from providing reliable transmission, carefully designed codes are also capable of securing the transmitted messages against an eavesdropper. On the other hand, it seems intriguing to design these codes because they are required to meet two conflicting requirements: $1$) they tend to add more redundancy to the codewords in order to counter the randomness of the channel and provide reliable communication, and $2$) in order to offer a secure transmission, they need to maximize the randomness perceived at the eavesdropper by reducing the redundancy. 

An important step towards designing codes for the wiretap channel is choosing a metric for measuring security. In contrast to measuring reliability where probability of error is being used as a universally accepted metric, there are multiple metrics for quantifying secrecy for this scenario as listed in Table \ref{SecMet}. Computation of these metrics for explicit finite length coding schemes, however, has been limited to some specific channels\footnote{The authors in \cite{explicit} provide a coding scheme for an AWGN channel, which satisfy strong secrecy using capacity approaching codes of infinite length and hash functions.}. One example is \cite{survey} in which the authors apply semantic secrecy to a finite length Reed-Muller code used over a binary erasure channel (BEC). For AWGN channels, a BER based metric called security gap (as mentioned in Section \ref{Error Probability/SINR Based Metrics}) is widely used by the research community, which is easily applicable to all practical coding schemes, however, its relation to the information theoretic notions is not established in a rigorous manner.

Security gap is defined as the difference between the qualities (SNRs) of the main and the eavesdropper's channels. For AWGN channels without feedback, this value must be greater than zero in order to result in a positive secrecy capacity (which equals to the difference between the capacities of the main and the eavesdropper's channels\cite{Leung}). Small security gaps are desirable because they make physical layer security achievable even with a small degradation of the eavesdropper's channel with respect to the main one. An analytical definition of the security gap can be given as follows. Let us fix two threshold values for the maximum desired BER over the main channel (denoted by $P_{B}^{max} \leq \frac{1}{2}$) and and the minimum desired BER at the eavesdropper (denoted by $P_{E}^{min}>0$). Then, the security gap equals the difference between the SNR values corresponding to these two, i.e., $SNR_{B}-SNR_{E}$ measured in dBs. 

The major shortcoming of the above BER-based analysis is that it assumes that if $P_{eve}^{min}\approx0.5$, then Eve cannot extract any information about the secret message. However, if there exists any statistical correlation between the bit errors or between the Eve's observation and the transmitted message, the amount of leaked information will not be zero. The authors in \cite{BERbased} employ tools from cryptography literature to eliminate such correlations. Specifically, they propose the use of substitution permutation networks (SPNs) without any secret key to satisfy the strict avalanche criterion (SAC) and the bit independence criterion (BIC). The required SPNs can be implemented efficiently \cite{BERbased}, hence, they can be applied to the message bits before the encoding stage. Therefore, one can design practical coding schemes, which result in BER very close to $0.5$ with no (or, small) correlation at Eve.

\begin{figure}[ht!]
	\includegraphics[width=70mm]{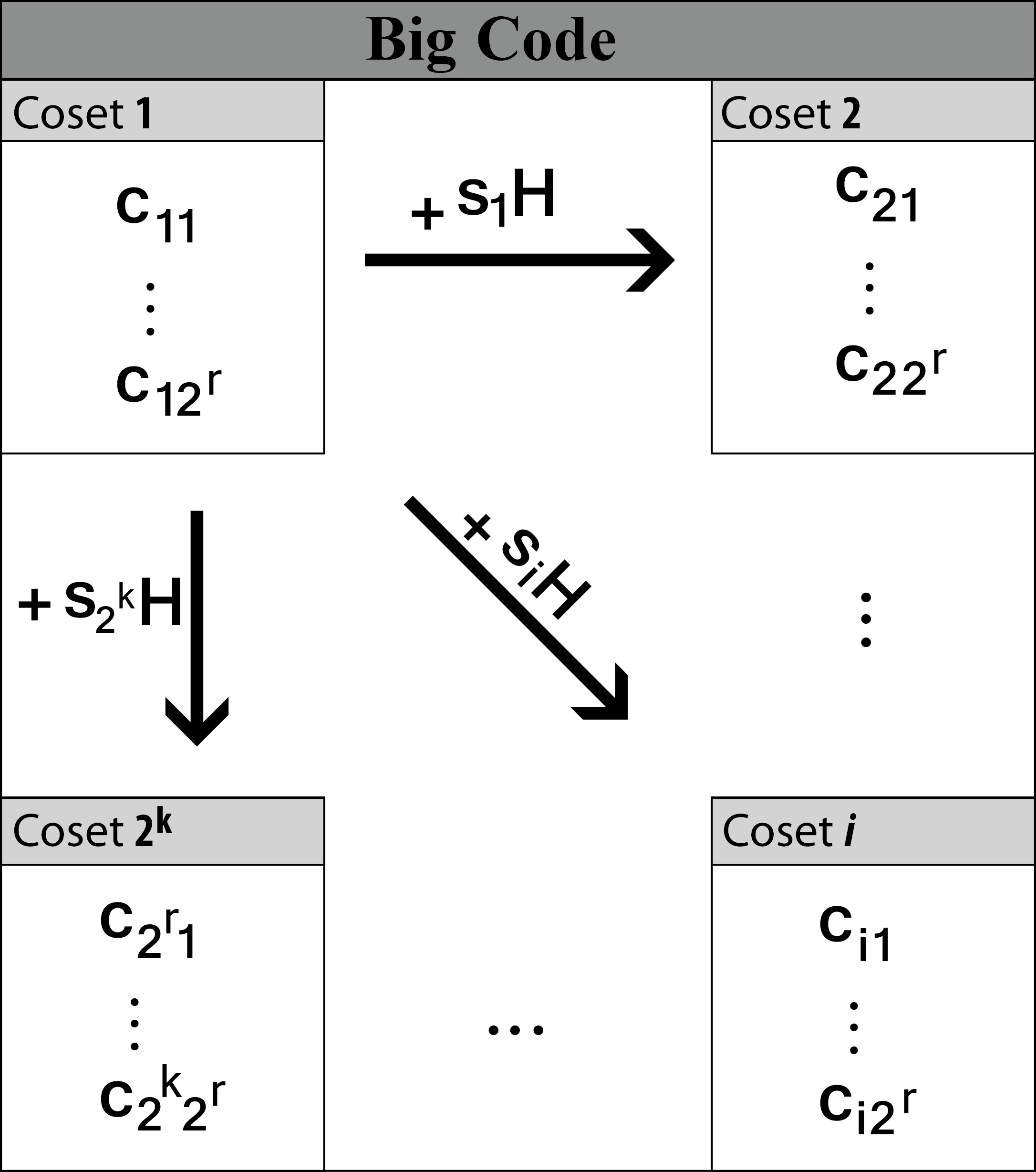}
	\centering
	\caption{Illustration of the coset coding scheme.}
	\label{fig:cosetcoding}
\end{figure}

The general coding scheme for physical layer security is the randomized encoding proposed by Wyner in \cite{Wyn74} to prove that there exists a code that achieves the secrecy capacity of the degraded wiretap channel. This method is also known as \textit{coset-coding} studied further in the subsequent literature, e.g., in \cite{ozarow} and \cite{syndrome}. In this method, all-zero message is mapped to a code $\mathcal{C}$, with generator matrix $\mathbf G$, and a non-zero message $\mathbf s=[s_1,s_2,...,s_k]$ is mapped to the coset obtained by adding the $n$-tuple $s_1\mathbf h_1+s_2\mathbf h_2+...+s_k\mathbf h_k$ to all the codewords in $\mathcal{C}$ where $\mathbf h_i$'s are linearly independent $n$-tuples outside $\mathcal{C}$. Assuming $\mathbf h_i$'s form a matrix $\mathbf H$ of size $k\times n$, the transmitted codeword through the channel is 
\begin{equation}
\label{eq:randomencoding}
\mathbf c=\begin{bmatrix}
\mathbf s&
\mathbf v
\end{bmatrix}
\begin{bmatrix}
\mathbf H \\
\mathbf G
\end{bmatrix}
\end{equation}
where $\mathbf v=[v_1,v_2,...,v_r]$ denotes the random bit vector used for choosing a codeword in the corresponding coset uniformly randomly. Fig. \ref{fig:cosetcoding} illustrates this method where \textit{small code} is denoted by $\mathcal{C}$ and $\mathbf c_{ij}$ denotes the $i^{\text{th}}$ codeword in the $j^{\text{th}}$ coset. We note that in conventional encoding, the leakage of information about the secret message to Eve, $H(\mathbf S^K|\mathbf Z^n)$, is equal to the leakage of the codewords $H(\mathbf c^n|\mathbf Z^n)$ \cite{survey} since there is a one-to-one mapping between the messages and the codewords. In the randomized encoding scheme, however, this is not true anymore since different codewords may represent the same message.

Application of lattice codes to the Gaussian wiretap channel is studied in \cite{lattice} where the authors define an alternative security metric (called secrecy gain), which is related to the theta series of lattices and shows the amount of confusion at the eavesdropper. Without introducing a decoding method, they evaluate the performance of different lattices based on the secrecy gain. The confusion at the eavesdropper in \cite{lattice} is the result of using a random lattice in addition to the lattice that is responsible for transmitting the original message. As a further development, the authors in \cite{lattice2} prove the existence of lattices that are semantically secure in the Gaussian wiretap channel. 

The authors in \cite{lowerbound1} and \cite{lowerbound2} provide a lower bound on the mutual information between the secret message and Eve's observation. They utilize this lower bound to quantify the equivocation for finite length codes for Gaussian wiretap channels. Specifically, they use finite length LDPC codes and demonstrate the achievable points in the equivocation-rate region. In \cite{BALDI2017}, the level of security at the physical layer is assessed from the information theoretic standpoint while taking into account the constraints of practical transmissions over realistic wireless wiretap channels, i.e., by considering practical codes with finite length, discrete modulation formats and continuous channels with fading.

\begin{table}[t!]
	\centering
	\caption{Practical coding schemes aiming at reducing the security gap over AWGN channel.}
	\label{tbl:PracticalCoding}
	\begin{tabular}{|p{3.5cm}|p{4cm}|p{7cm}|}
		\hline
		Paper  & Secure coding scheme & Contributions \\ \hline
		D. Klinc \textit{et al.} \cite{puncture} &  Punctured LDPC codes & Propose puncturing information bits in LDPC codes in order to confuse Eve.\\ \hline
		M. Baldi \textit{et al.} \cite{scramble} & Non-systematic codes                            & Make use of scrambling technique at encoding which amplifies the resulting errors at decoding stage. \\ \hline
		M. Baldi \textit{et al.} \cite{scramblesecond} & Coding with scrambling, concatenation, and HARQ & Investigate different techniques to reduce the resulting security gaps.\\ \hline
		Y. X. Zhang \textit{et al.} \cite{polar-LDPC} & Polar-LDPC concatenated codes & Evaluate performance of a serially concatenated coding scheme consists of  polar and LDPC codes. \\ \hline
		A. Nooraiepour and T. M. Duman \cite{RandomizedConvolutional}  & Randomized convolutional and turbo codes        & Construct coset codes based on convolutional and turbo codes using dual codes. \\ \hline
	\end{tabular}
\end{table}

\begin{figure}[t!]
	\centering
	\includegraphics[width=13cm]{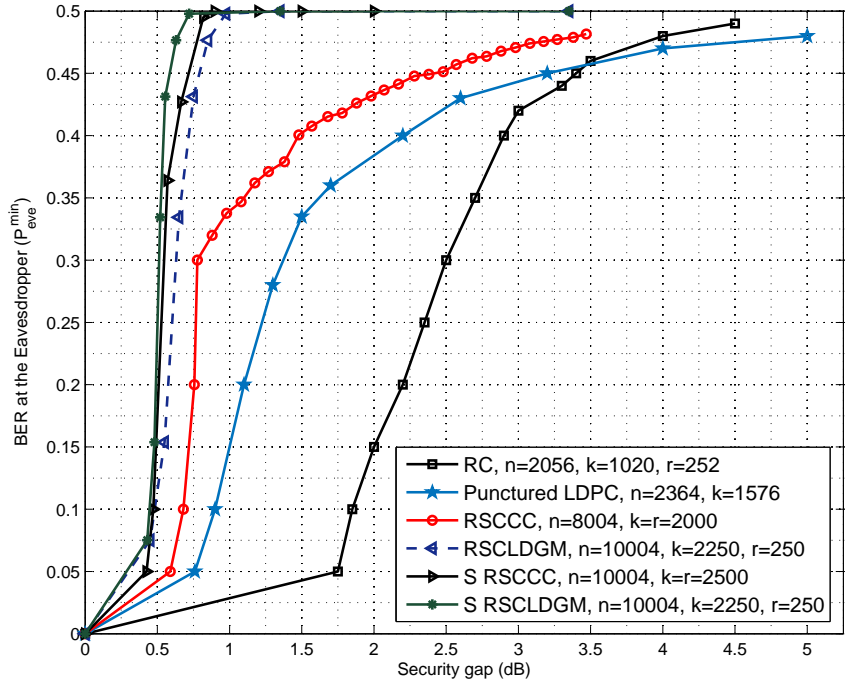}
	\caption{The resulting security gaps from practical coding schemes. $n$, $k$ and $r$ denote length of a code, number of data bits and number of random bits, respectively. S means scrambling has been applied. Taken from \cite{RandomizedConvolutional} and \cite{RandomizedSCLDGM}.}
	\label{fig:secgaps}
\end{figure}

Several practical coding schemes aiming directly at reducing the security gap are also proposed in the literature. Specifically, punctured LDPC codes are exploited for physical layer security in \cite{puncture}. Furthermore, \cite{scramble} demonstrates how non-systematic codes can be effective to reduce the resulting security gap, while \cite{scramblesecond} applies different techniques including scrambling, concatenation, and hybrid automatic repeat-request to LDPC and BCH codes in order to further reduce the security gap. 
Concatenation of polar and LDPC codes are proposed in \cite{polar-LDPC}. Coset codes based on convolutional and turbo codes are constructed in \cite{RandomizedConvolutional} and \cite{RandomizedTurbo} for physical layer security. Moreover, application of serially concatenated low density generator matrix (SCLDGM) codes to the randomized scheme is studied in \cite{RandomizedSCLDGM}. 

Authors in \cite{explicit} introduce a method, which provides strong secrecy for the Gaussian wiretap channel by means of hash functions and capacity approaching channel codes of infinite length. Using this scheme, it is also possible to quantify the level of leakage obtained for a fixed block length. Finally, \cite{finite} provides the maximal secrecy rate over a wiretap channel subject to reliability and secrecy constraints at a given block length.

Table \ref{tbl:PracticalCoding} summarizes the existing practical coding schemes for Gaussian wiretap channel along with their main contributions. Furthermore, Fig. \ref{fig:secgaps} demonstrates the relation between the resulting security gaps from some of these practical coding schemes\footnote{RSCCC stands for randomized serially concatenated convolutional codes scheme, RC stands for randomized convolutional codes scheme and RSCLDGM corresponds to the randomized SCLDGM codes.} and a specific BER at Eve. One can see that when a BER of $0.5$ is desired at Eve, security gaps as small as $1$ dB can be obtained using carefully designed codes of length $10^4$ or longer.

In addition to the Gaussian wiretap channels, some efforts are also made to develop practical coding schemes for fading wiretap channels. The authors in \cite{Zheng} and \cite{Si} propose polar coding schemes for achieving secrecy over block fading wiretap channels. In \cite{Zheng}, the secure coding scheme is designed with the aid of instantaneous MCSI and ECSI. However, the polar coding approach proposed in \cite{Si} relies on statistical CSI of both channels. Application of algebraic lattice codes is also proposed over block fading wiretap channels. For instance, a sequence of non-random lattice codes is developed in \cite{Luzzi} that achieve strong secrecy and semantic security over ergodic fading channels.

We emphasize that the easy-to-compute property of security gap makes it a widely used metric in majority of the aforementioned code designs, which try to avoid potential complexities associated with computing the original information-theoretic metrics for finite-length scenarios. Despite some limited efforts such as \cite{lowerbound2} in providing a lower bound for information-theoretic metrics (Table \ref{SecMet}), the literature of physical layer security still lacks a comprehensive and satisfactory finite length analysis of different metrics especially for AWGN and fading wiretap channels.

\section{Lessons Learned and the Challenges Ahead}
\label{sec:challenges}

We now provide a summary of the lessons learned and discuss a number of challenges ahead and present some directions for future research in this area.



\subsection{Lessons Learned}

The usual approaches for designing secure transmit signals with Gaussian inputs lead to considerable losses when applied to the signals drawn from finite constellations. This fact motivates the need for development of secure transmission schemes under the finite-alphabet input assumption.

In single-antenna transmission with constellation constrained inputs, achieving secrecy requires adoption of a proper power allocation strategy. This is because transmission with full power may allow the eavesdropper to successfully detect the transmitted symbols. Moreover, the increased dimensionality in multi-antenna transmissions can be employed to enhance secrecy under finite-alphabet inputs. Precoding serves as a promising strategy for maximizing the quality difference between the signals received at the legitimate receiver and the eavesdropper. Artificial noise injection offers considerable secrecy gains and it well-suits the nature of secure transmission with finite-alphabet inputs since the optimal power, which is allocated to the information bearing signal is only a fraction of the total power, and the excess power can be used for artificial noise injection. In many cases (especially, in the absence of instantaneous ECSI), without artificial noise injection, secrecy rates drop to zero when the transmit power increases. Therefore, injecting artificial noise is crucial for guaranteeing a good saturation behavior at high SNRs. 

The optimal secure transmission schemes are not known in closed-form in almost all of the scenarios studied in this survey. This is due to the non-convexity of the objective function (i.e., secrecy rate). Furthermore, the transmit signal design algorithms, which rely on direct maximization of secrecy rates, are computationally complex mainly due to the fact that the mutual information expression under finite-alphabet inputs lacks closed-form. Therefore, different studies focus their attention on proposing lower-complexity alternatives to direct secrecy rate maximization. There is also an important trade-off between the secrecy rates and the computational complexity of the existing solutions in the literature of physical layer security with finite-alphabet inputs. Moreover, it is noteworthy that the transmit signal designs based on metrics such as SINR or BER, though simple and may be practical in some scenarios, do not ensure secrecy in an information theoretic sense.
	
The amount of CSI at the transmitter plays a central role in determining the secrecy performance of a particular transmission scheme. Transmission algorithms employing full CSI typically have a high capability in simultaneously increasing the information rates at the legitimate receiver and suppressing the reception at the eavesdropper, however, acquiring a perfect instantaneous CSI of the eavesdropper may not be practically possible in many scenarios. The achievable secrecy rates undergo considerable losses in the absence of instantaneous ECSI. However, nonzero secrecy rates are still attainable with the aid of statistical ECSI or in some cases even without it (i.e., with the aid of the MCSI only).

Secure transmission design over multiuser channels, e.g., broadcast channel, is a more challenging task compared to the point-to-point scenarios. This is because the proposed solutions need to meet additional requirements e.g., in terms of inter-user interference. Similarly, various challenges arise when moving to the scenarios with multiple eavesdroppers. For instance, secure transmission along the null-space of the eavesdropper's channel is no longer possible in these scenarios and designing an artificial noise signal which effectively degrades the reception at the eavesdroppers is more complex than the single-eavesdropper case.

Practical coding schemes exploit different tools from coding theory and cryptography in order to ensure a certain level of security and reliability (in terms of BERs) for Eve and Bob, respectively. By combining coset coding and scrambling approaches along with code concatenation, security gaps as small as $1$ dB can be obtained using finite-length explicit and implementable codes. However, finding specific (and implementable)  finite-length codes, which can provide physical layer security from an information-theoretic point of view, remains as a future challenge.

\subsection{Challenges and Open Problems}

Various studies have been carried out on physical layer security with discrete signaling with the objective of bridging the gap between the information theoretic limits of secrecy and practical and implementable secure transmission schemes. However, there still are many important open problems and future challenges to be addressed in this area.

%

Physical layer security with finite-alphabet inputs has been studied only for a limited set of CSI scenarios including the cases with perfect and statistical MCSI and ECSI at the transmitter. Some other important scenarios regarding the CSI uncertainty at different nodes, e.g., noisy estimation of CSI, outdated CSI or limited CSI feedback, which are well-investigated under the Gaussian input assumption \cite{Hyadi}, have not yet been considered for cases with finite-alphabet inputs. It would be interesting and of significant practical value to quantify the losses in the secrecy rates under these different imperfect CSI scenarios in the context of finite-alphabet inputs.

While the majority of artificial noise injection strategies rely on transmitting Gaussian distributed noise either by aligning it in the null-space of the main channel or by optimizing its covariance matrix, such artificial noise signals are not necessarily optimal. For instance, it has been shown in \cite{Shamai_Verdu} that the worst case noise for AWGN channels with binary inputs has a discrete distribution. This motivates seeking interference distributions (rather than merely optimizing covariance matrices of Gaussian distributed signals), which maximize secrecy rates. 


While some steps have been taken in studying physical layer security over finite-input multiuser networks (as surveyed in Section \ref{sec:multiuser}), many important scenarios have not yet been tackled. For example, no results have been reported for interference channels with secrecy constraints under the finite-alphabet input assumption (for an example with Gaussian inputs refer to \cite{Ayca}). Secrecy issues concerning the networks with simultaneous wireless information and power transfer (SWIPT) is also a topic of recent interest \cite{Chen}. It is demonstrated that SWIPT networks are prone to information leakage at the energy receivers, and hence, the main focus of the existing articles is the characterization of the trade-off between the secrecy capacity/rate and the harvested power in these networks \cite{R_Zhang13}-\hspace{-0.2px}\cite{Ulukus16}. Studying this trade-off under practical discrete signaling assumption is an important subject for future research. Furthermore, physical layer security in heterogeneous networks \cite{Wang} and with full duplex transmitter or receivers (see, e.g. \cite{YZhou} and \cite{Abedi}) are other hot topics in this area, and extending the existing results for these scenarios to the case of finite-alphabet inputs constitutes another important direction for future research. 

Massive MIMO systems provide considerable gains in secrecy capacity. Besides opportunities offered, many challenges arise in secure transmit signal designs with finite-alphabet inputs. For instance, most of the existing precoder design approaches (such as the ones proposed in \cite{Wu} and \cite{TWC}) are intractable when the number of antennas is high. Secure mmWave transmission is another promising research frontier. Communicating at the mmWave spectrum considerably changes the channel model with respect to the microwave communication systems, and introduces some new features (e.g., high sensitivity to blockages and considerable differences between the line-of-sight and non-line-of-sight receptions), which motivates development of new frameworks for designing practical secure transmission schemes.

Taking into account the hardware constraints in designing secure massive MIMO design is another topic worth investigating. For example, secure transmission schemes can be proposed with finite-alphabet inputs for multi-antenna transmitters with a single radio frequency (RF) chain (or a few, but less than the number of antennas). Performance of wireless systems can be considerably affected by different RF impairments such as impairments such as inphase/quadrature-phase (I/Q) imbalance, phase noise or distortions due to the nonlinear power amplifiers, which degrade the reception at the receivers (including the eavesdropper); hence, this behavior can be exploited to enhance secrecy. Moreover, various practical constraints arise in IoT systems (e.g., in terms of complexity and energy efficiency) that should be considered in designing secure transmission schemes. We also note that, designing testbeds and conducting tests by implementing physical layer security schemes is an essential step, which should be taken before these techniques can be adopted by actual communication systems. One example of such experiments is reported in \cite{Ogg} where the authors evaluate the secrecy performance of coset coding of lattice constellations with the aid of a software defined radio testbed.


Another interesting direction for future work is to further develop cross-layer design based solutions for securing communications as only a limited number of works have been reported on this topic so far. For instance, the authors in \cite{Ham_cross} demonstrate the advantages of cross-layer protocol interactions on the achievable secrecy by joint exploitation of automatic-repeat-request (as a MAC layer operation) and maximal ratio combining (as a physical layer operation). The potentials of other layers can also be taken advantage of to enhance security. For instance, employing authentication and watermarking strategies at the application layer along with the coding and signal processing at the physical layer \cite{LZhou14} can be lead to considerable secrecy gains.

The main problem associated with designing practical coding schemes for physical layer security is the following: Although the security gap is proposed and used as an alternative practical metric, the most widely accepted one (i.e., equivocation) relies on information theoretic quantities, and computing these quantities for finite-length codes over AWGN channels is a challenging problem, and can be a promising research direction on physical layer security. As a final suggestion for future research, we recommend the problem of precoder design in the scenarios where practical wiretap codes are employed, which would be an important step towards designing implementable secure transmission schemes.


\section{Conclusions}
\label{sec:Conclusion}
An obligatory step towards practical implementation of physical layer security is understanding the impacts of employing standard constellations on the secrecy performance. With this motivation, this survey paper provides a summary of the current literature on physical layer security with discrete signaling and practical codes. We outline why and how transmit signal design problem is different for the case of finite-alphabet inputs compared to the case of Gaussian signals, and review existing algorithms for maximizing secrecy rates in a variety of scenarios. We also provide an overview on practical code constructions for physical layer security, and point out important challenges and open problems on this topic.

%
%
%

\section*{Acknowledgment}

This work was supported by the Scientific and Technical Research Council of Turkey (TUBITAK) under grant \#113E223.


\begin{thebibliography}{136}
	
	
	\bibitem{Liang}  Y. Liang, H. V. Poor, and S. Shamai, \textit{Information Theoretic Security}, Found. Trends Commun. Inf. Theory, vol. 5, no. 4-5, pp. 355-580, 2008.
	
	\bibitem{Liu} R. Liu and W. Trappe, \textit{Securing Wireless Communications at the Physical Layer}, Springer, 2010.
	
	\bibitem{Jorswieck10} E. A. Jorswieck, A. Wolf, and S. Gerbracht, \textquotedblleft Trends in telecommunications technologies," \textit{in Secrecy on the Physical Layer in Wireless Networks.} Croatia, InTech, 2010.
	
	\bibitem{Bloch_book} M. Bloch and J. Barros, \textit{Physical-layer security: from information theory to security engineering}, Cambridge, 2011.
	
	\bibitem{Guvenkaya17} E. Guvenkaya, J. M. Hamamreh, and H. Arslan, \textquotedblleft On physical-layer concepts and metrics in secure signal transmission," \textit{Phys. Commun.}, vol. 25, pp. 14–25, Dec. 2017.
	
	\bibitem{Shiu11} Y.-S. Shiu, S. Y. Chang, H.-C. Wu, S. C.-H. Huang, and H.-H. Chen, \textquotedblleft Physical layer security in wireless networks: A tutorial," \textit{IEEE Wireless Commun.}, vol. 18, no. 2, pp. 66–74, Apr. 2011.
	
	\bibitem{Mukherjee} A. Mukherjee, S. A. A. Fakoorian, J. Huang, and A. L. Swindlehurst, \textquotedblleft Principles of physical layer security in multiuser wireless networks: A survey," \textit{IEEE Commun. Surveys Tuts.}, vol. 16, no. 3, pp. 1550--1573, Aug. 2014.
	
	\bibitem{Hyadi} A. Hyadi, Z. Rezki and M. S. Alouini, \textquotedblleft An overview of physical layer security in wireless communication systems with CSIT uncertainty," in \textit{IEEE Access}, vol. 4, pp. 6121--6132, Oct. 2016.
	
	\bibitem{Chen17} X. Chen, D. W. K. Ng, W. H. Gerstacker and H. H. Chen, \textquotedblleft A survey on multiple-antenna techniques for physical layer security," in \textit{IEEE Commun. Surveys Tuts.}, vol. 19, no. 2, pp. 1027--1053, Second Quarter 2017.
	
	\bibitem{NYang15} N. Yang, L. Wang, G. Geraci, M. Elkashlan, J. Yuan, and M. D. Renzo, \textquotedblleft Safeguarding 5G wireless communication networks using physical layer security," \textit{IEEE Commun. Mag.}, vol. 53, pp. 20–27, Apr. 2015.
	
	\bibitem{Wu_phy_sec} Y. Wu, A. Khisti, C. Xiao, G. Caire, K. K. Wong and X. Gao, \textquotedblleft A survey of physical layer security techniques for 5G wireless networks and challenges ahead," in \textit{IEEE J. Sel. Areas Commun.}, vol. 36, no. 4, pp. 679-695, Apr. 2018.
	
	\bibitem{YLiu_survey} Y. Liu, H. H. Chen, and L. Wang, \textquotedblleft Physical layer security for next generation wireless networks: Theories, technologies, and challenges," \textit{IEEE Commun. Surveys Tuts}, vol. 19, no. 1, pp. 347-376, Jan. 2017
	
	\bibitem{Trappe} W. Trappe, \textquotedblleft The challenges facing physical layer security," \textit{IEEE Commun. Mag.}, vol. 53, no. 6, pp. 16--20, Jun. 2015.
	
	\bibitem{semantic}
	M. Bellare, S. Tessaro, and A. Vardy, \textquotedblleft Semantic security for the wiretap channel,” in \textit{Proc. 32nd Annu. Cryptol. Conf.}, vol. 7417. 2012, pp. 294–311.	
	
	\bibitem{Wyn74}
	A. D. Wyner, \textquotedblleft The wire-tap channel," \textit{Bell Sys. Tech. J.}, vol. 54, pp. 1355--1387, 1975.
	
	\bibitem{Shannon} C. E. Shannon, \textquotedblleft Communication theory of secrecy systems," \textit{Bell Sys. Tech. J.}, vol. 28, pp. 656--715, 1949.
	
	\bibitem{Maurer} U. M. Maurer and S. Wolf, \textquotedblleft From weak to strong information-theoretic key agreement," in \textit{IEEE Int’l. Symp. Info. Theory (ISIT)}, p. 18, Sorrento, Italy, Jun. 2000.
	
	\bibitem{Leung}
	S. L. Y. Cheong and M. Hellman, \textquotedblleft The Gaussian wire-tap channel,'' \textit{IEEE Trans. Inf. Theory}, vol. 24, no. 4, pp. 451--456, Jul. 1978.
	
	\bibitem{Csiszar}
	I. Csiszar and J. Korner, \textquotedblleft Broadcast channels with confidential messages,'' \textit{IEEE Trans. Inf. Theory}, vol. 24, no. 3, pp. 339--348, May 1978.
	
	\bibitem{Barros06} 
	J. Barros and M. R. D. Rodrigues, \textquotedblleft Secrecy Capacity of Wireless Channels,” \textit{IEEE Int. Symp. Info. Theory (ISIT)}, Jul. 2006, pp. 356--360.
	
	\bibitem{Liang_BCC} Y. Liang, H. V. Poor, and S. Shamai, \textquotedblleft Secure communication over fading channels,” \textit{IEEE Trans. Inf. Theory}, vol. 54, no. 6, pp. 2470-2492, Jun. 2008. 
	
	\bibitem{Gopala} P. Gopala, L. Lai, and H. El Gamal, \textquotedblleft On the secrecy capacity of fading
	channels”, \textit{IEEE Trans. Inf. Theory}, vol. 54, no. 10, pp. 4687--4698, Oct. 2008.
	
	\bibitem{Lin16} P.-H. Lin and E. Jorswieck, \textquotedblleft On the fast fading Gaussian wiretap channel with statistical channel state information at the transmitter," \textit{IEEE Trans. Inf. Forensics Security}, vol. 11, no. 1, pp. 46--58, Jan. 2016.
	
	\bibitem{Khisti}
	A. Khisti and G. Wornell, \textquotedblleft Secure transmission with multiple antennas II: the MIMOME wiretap channel," \textit{IEEE Trans. Inf. Theory}, vol. 56, no. 11, pp. 5515--5532, Nov. 2010.
	
	\bibitem{Hassibi}
	F.~ Oggier, B.~ Hassibi, \textquotedblleft The secrecy capacity of the MIMO wiretap channel,'' {\it IEEE  Trans. Inf. Theory}, vol.~57,  no. 8, pp.~4961-4972,  2011.
	
	\bibitem{Shamai}
	T. Liu and S. Shamai, \textquotedblleft A note on the secrecy capacity of the multiple antenna wiretap channel," \textit{IEEE Trans. Inf. Theory}, vol. 55, no. 6, pp. 2547--2553, Jun. 2009.
	
	\bibitem{QLi} Q. Li, M. Hong, H. T. Wai, Y. F. Liu, W. K. Ma and Z. Q. Luo, \textquotedblleft Transmit solutions for MIMO wiretap channels using alternating optimization," in \textit{IEEE J. Sel. Areas Commun.}, vol. 31, no. 9, pp. 1714-1727, Sep. 2013.
	
	\bibitem{Loyka} S. Loyka and C. D. Charalambous, \textquotedblleft An algorithm for global maximization of secrecy rates in Gaussian MIMO wiretap channels," in \textit{IEEE Trans. Commun.}, vol. 63, no. 6, pp. 2288-2299, Jun. 2015.
	
	\bibitem{JLi} J. Li and A. P. Petropulu, \textquotedblleft On ergodic secrecy rate for Gaussian MISO wiretap channels," \textit{IEEE Trans. Wireless Commun.}, vol. 10, no. 4, pp. 1176-1187, Apr. 2011.
	
	\bibitem{Klinc} D. Klinc, J. Ha, S. W. McLaughlin, J. Barros, and B.-J. Kwak, \textquotedblleft LDPC codes for the Gaussian wiretap channel,” \textit{IEEE Trans. Inf. Forensics Security}, vol. 6, no. 3, pp. 532–540, Sep. 2011.
	
	\bibitem{Liao} W.-C. Liao, T.-H. Chang, W.-K. Ma, and C.-Y. Chi, \textquotedblleft QoS-based transmit beamforming in the presence of eavesdroppers: An optimized artificial-noise aided approach,” \textit{IEEE Trans. Signal Processing}, vol. 59, no. 3, pp. 1202–1216, Mar. 2011.
	
	\bibitem{Raghava} G. D. Raghava and B. S. Rajan, \textquotedblleft Secrecy capacity of the Gaussian wiretap channel with finite complex constellation input," [Online]. Available: http://arxiv.org/abs/1010.1163, Oct. 2010.
	
	\bibitem{Rodrigues} M. R. D. Rodrigues, A. Somekh-Baruch, and M. Bloch, \textquotedblleft On Gaussian wiretap channels with M-PAM inputs," \textit{in Proc. EWConf.}, Apr. 2010, pp. 774--781.
	
	\bibitem{Biglieri} E. Biglieri, \textit{Coding for wireless channels}, Springer-Verlag New York, 2005.
	
	\bibitem{ChaoQi} C. Qi, Y. Chen, and A. J. Vinck. \textquotedblleft On the Binary Input Gaussian Wiretap Channel with/without Output Quantization," \textit{Entropy}, vol. 19, no. 2, Feb. 2017.
	
	\bibitem{ZLi} Z. Li, R. Yates and W. Trappe, \textquotedblleft Achieving secret communication for fast Rayleigh fading channels," in \textit{IEEE Trans. Wireless Commun.}, vol. 9, no. 9, pp. 2792-2799, Sep. 2010.
	
	
	\bibitem{Goel}
	S. Goel and R. Negi, \textquotedblleft Guaranteeing secrecy using artificial noise," \textit{IEEE Trans. Wireless Commun.}, vol. 7, no. 6, pp. 2180--2189, Jun. 2008.
	
	\bibitem{Bashar}
	S. Bashar, Z. Ding, and C. Xiao, \textquotedblleft On secrecy rate analysis of MIMO wiretap channels driven by finite-alphabet input," \textit{IEEE Trans. Commun.}, vol. 60, no. 12, pp. 3816--3825, Dec. 2012.
	
	\bibitem{Wu} 
	Y. Wu, C. Xiao, Z. Ding, X. Gao, and S. Jin, \textquotedblleft Linear precoding for finite-alphabet signaling over MIMOME wiretap channels," \textit{IEEE Trans. Veh. Technol.}, vol. 61, no. 6, pp. 2599--2612, Jul. 2012.
	
	\bibitem{Khisti_MISOSE} A. Khisti and G. W. Wornell, \textquotedblleft Secure transmission with multiple antennas—Part I: The MISOME wiretap channel," \textit{IEEE Trans. Inf. Theory}, vol. 56, no. 7, pp. 3088--3104, Jul. 2010.
	
	\bibitem{Khandaker} M. R. A. Khandaker, C. Masouros, K. Wong, \textquotedblleft Constructive interference based secure precoding: a new dimension in physical layer security," \textit{IEEE Trans. Inf. Forensics Security}, vol. 13, no. 9, pp. 2256-2268, Sep. 2018.
	
	
	\bibitem{Reboredo} H. Reboredo, J. Xavier, and M. R. D. Rodrigues, \textquotedblleft Filter design with secrecy constrains: The MIMO Gaussian wiretap channel,\textquotedblleft \textit{IEEE Trans. Signal Process.}, vol. 61, pp. 3799–3814, Aug. 2013.
	
	%
	
	\bibitem{Nguyen}
	T. V. Nguyen, T. Q. S. Quek, Y. H. Kim, and H. Shin, \textquotedblleft Secrecy diversity in MISOME wiretap channels," in \textit{Proc. Global Commun. Conf. (Globecom’2012)}, Aneheim, USA, Dec. 2012, pp. 4840–4845.
	
	\bibitem{Bashar0}
	S. Bashar, Z. Ding, and C. Xiao, \textquotedblleft On the secrecy rate of multi-antenna wiretap channel under finite-alphabet input," \textit{IEEE Commun. Lett.}, vol. 15, no. 5, pp. 527-–529, May 2011.
	
	\bibitem{ISIT} S. Rezaei Aghdam and T. M. Duman, \textquotedblleft Low complexity precoding for MIMOME wiretap channels based on cut-off rate," \textit{IEEE Int. Symp. Inf. Theory (ISIT 2016)}, Barcelona, Jul. 2016, pp. 2988--2992.
	
	\bibitem{GAN} P.-H. Lin S.-H. Lai, S.-C. Lin, and H.-J. Su, \textquotedblleft On secrecy rate of the generalized artificial-noise assisted secure beamforming for wiretap channels,” \textit{IEEE J. Sel. Areas Commun.}, vol. 31, no. 9, pp. 1728-1740, Sep. 2013.
	
	\bibitem{TWC} S. Rezaei Aghdam, T. M. Duman, \textquotedblleft Joint precoder and artificial noise design for MIMO wiretap channels with finite-alphabet inputs based on the cut-off rate," \textit{IEEE Trans. Wireless Commun.}, vol. 16, no. 6, pp. 3913-3923, Jun. 2017.
	
	\bibitem{Kalantari_conf} A. Kalantari, M. Soltanalian, S. Maleki, S. Chatzinotas, and B. Ottersten, \textquotedblleft Secure M-PSK communication via directional modulation," in \textit{Proc. IEEE ICASSP}, Shanghai, China, Mar. 2016, pp. 3481-3485
	
	\bibitem{Kalantari_jour} A. Kalantari, M. Soltanalian, S. Maleki, S. Chatzinotas and B. Ottersten, \textquotedblleft Directional modulation via symbol-level precoding: A way to enhance security," in \textit{IEEE J. Sel. Topics Signal Process.}, vol. 10, no. 8, pp. 1478-1493, Dec. 2016.
	
	\bibitem{THP} L. Zhang, Y. Cai, B. Champagne, and M. Zhao, \textquotedblleft Tomlinson-Harashima precoding design in MIMO wiretap channels based on the MMSE criterion,” in \textit{Proc. IEEE ICC}, pp. 470-474, Jun. 2015.
	
	\bibitem{cyclostationary}
	A. Punchihewa, Q. Zhang, O. A. Dobre, C. Spooner, S. Rajan and R. Inkol, \textquotedblleft On the cyclostationarity of OFDM and single carrier linearly digitally modulated signals in time dispersive channels: theoretical developments and application," in \textit{IEEE Trans. Wireless Commun.}, vol. 9, no. 8, pp. 2588-2599, Aug. 2010.
	
	\bibitem{Fakoorian} S. A. A. Fakoorian, H. Jafarkhani, and A. L. Swindlehurst, \textquotedblleft Secure space-time block coding via artificial noise alignment,” in \textit{Proc. Conf. Rec. 45th ASILOMAR (ASILOMAR’2011)}, Pacific Grove, USA, Nov. 2011, pp. 651–655.
	
	\bibitem{Hamamreh2016} J. M. Hamamreh, E. Guvenkaya, T. Baykas and H. Arslan, \textquotedblleft A practical physical-layer security method for precoded OSTBC-based systems," \textit{2016 IEEE Wireless Commun. Netw. Conf. (WCNC)}, Doha, 2016, pp. 1-6.
	
	
	\bibitem{Girnyk_ISITA} M. A. Girnyk, M. Vehkapera, J. Yuan, and L. K. Rasmussen, \textquotedblleft On the ergodic secrecy capacity of MIMO wiretap channels with statistical CSI,” in \textit{Proc. Int. Symp. Inf. Theory Appl. (ISITA)}, Melbourne, VIC, Australia, Oct. 2014, pp. 398–402.
	
	\bibitem{PIMRC17} S. Rezaei Aghdam, T. M. Duman, \textquotedblleft Transmit signal design for MIMO wiretap channels with statistical CSI and arbitrary inputs," in \textit{IEEE Int. Symp. Pers. Indoor Mobile Radio Commun. (PIMRC 2017)}, Montreal, QC, Oct. 2017, pp. 1-5.
	
	\bibitem{XLi} X. Li, R. Fan, X. Ma, J. An, and T. Jiang, \textquotedblleft Secure space-time communications over Rayleigh flat fading channels," \textit{IEEE Trans. Wireless Commun.}, vol. 15, pp. 1491–1504, Feb. 2016.
	
	\bibitem{Furqan} H. M. Furqan, J. M. Hamamreh and H. Arslan, \textquotedblleft Secret key generation using channel quantization with SVD for reciprocal MIMO channels," \textit{ISWCS}, Poznan, Sep. 2016, pp. 597–602.
	
	\bibitem{T_Allen_rec} T. Allen, J. Cheng and N. Al-Dhahir, \textquotedblleft Secure space-time block coding without transmitter CSI," \textit{IEEE Wireless Commun. Lett.}, vol. 3, no. 6, pp. 573-576, Dec. 2014.
	
    \bibitem{T_Dean} T. Dean and A. Goldsmith,  \textquotedblleft Physical-layer cryptography through massive MIMO," \textit{2013 IEEE Inf. Theory Workshop (ITW)}, Sevilla, 2013, pp. 1-5.
    
	
	
	\bibitem{Zeng} W. Zeng, C. Xiao, M. Wang, and J. Lu, \textquotedblleft Linear precoding for finite-alphabet inputs over MIMO fading channels with statistical CSI," \textit{IEEE Trans. Signal Process.}, vol. 60, no. 7, Jul. 2012.
	
	\bibitem{Girnyk_Spr} M. A. Girnyk, F. Gabry, M. Vehkapera, L. K. Rasmussen, and M. Skoglund, \textquotedblleft On the transmit beamforming for MIMO wiretap channels: Large-system analysis," in \textit{Proc. Int. Conf. Inf. Theoretic Secur. (ICITS)}, Singapore, Nov. 2013, pp. 90--102.
	
	\bibitem{Girnyk15} M. A. Girnyk, F. Gabry, M. Vehkapera, L. K. Rasmussen and M. Skoglund, \textquotedblleft MIMO wiretap channels with randomly located eavesdroppers: Large-system analysis," 2015 \textit{IEEE International Conference on Communication Workshop (ICCW)}, London, Jun. 2015, pp. 480-484.
	
	\bibitem{Wu17} Y. Wu, J. Wang, J. Wang, R. Schober, and C. Xiao, \textquotedblleft Large-scale MIMO secure transmission with finite
	alphabet inputs," [Online]. Available: https://arxiv.org/pdf/1612.08328.pdf, Dec. 2016.
	
	\bibitem{Mesleh} R. Mesleh, H. Haas, S. Sinanovic, C. W. Ahn, and S. Yun, \textquotedblleft Spatial modulation,” \textit{IEEE Trans. on Vehic. Technol.}, vol. 57, no. 4, pp. 2228–2241, 2008.
	
	\bibitem{Jeganathan} J. Jeganathan, A. Ghrayeb, L. Szeczecinski, A. Ceron, \textquotedblleft Space shift keying modulation for MIMO channels," \textit{IEEE Trans. Wireless Commun.} vol. 8, no. 7, pp. 3692-3703, Jul. 2009. 
	
	\bibitem{Prc13} M.~Di Renzo, H.~Haas, A.~Ghrayeb, S.~Suguira and L.~Hanzo \textquotedblleft Spatial modulation for generalized MIMO: challenges, opportunities and implementation," {\it Proc. IEEE}, vol.~102, no.~1,  pp.~56--103, 2014.
	
	\bibitem{Sinanovic} S. Sinanovic, N. Serafimovski, M. Di Renzo, and H. Haas, \textquotedblleft Secrecy capacity of space keying with two antennas,” in \textit{2012 IEEE  Veh. Technol. Conf. (VTC Fall)}, Sep. 2012, pp. 1–5.
	
	\bibitem{Guan} X. Guan, Y. Cai, and W. Yang, \textquotedblleft On the secrecy mutual information of spatial modulation with finite alphabet,” in \textit{Proc. IEEE Int. Conf. Wireless Commun. Signal Process. (WCSP)}, Oct. 2012, pp. 1–4
	
	\bibitem{Conf} S. Rezaei Aghdam, T. M. Duman and M. Di Renzo, \textquotedblleft On secrecy rate analysis of spatial modulation and space shift keying,", {\it IEEE BlackSeaCom 2015}, pp.~ 63-67, May 2015.
	
	\bibitem{Lett}
	S. Rezaei Aghdam and T. M. Duman, \textquotedblleft Physical layer security for space shift keying transmission with precoding," \textit{IEEE Wireless Commun. Lett.}, vol. 5, no. 2, pp. 180-183, Apr. 2016.
	
	\bibitem{SPSM} F. Wu, L. L. Yang, W. Wang and Z. Kong, \textquotedblleft Secret precoding-aided spatial modulation," in \textit{IEEE Commun. Lett.}, vol. 19, no. 9, pp. 1544-1547, Sep. 2015.
	
	\bibitem{Wang_SM} L. Wang, S. Bashar, Y. Wei, and R. Li, \textquotedblleft Secrecy enhancement analysis against unknown eavesdropping in spatial modulation," \textit{IEEE Commun. Lett.}, vol. 19, no. 8, pp. 1351-1354, Aug. 2015.
	
	\bibitem{QSM} Z. Huang, Z. Gao and L. Sun, \textquotedblleft Anti-eavesdropping scheme based on quadrature spatial modulation," in \textit{IEEE Commun. Lett.}, vol. 21, no. 3, pp. 532-535, Mar. 2017.
	
	
	%
	
	
	
	\bibitem{Li_SM} Q.-L. Li, \textquotedblleft Information-guided randomization for wireless physical layer secure transmission," in \textit{Proc. IEEE Military Commun. Conf.}, Orlando, FL, USA, Nov. 2012, pp. 1--6.
	
	\bibitem{Globecom}  S. Rezaei Aghdam and T. M. Duman, \textquotedblleft Secure space shift keying transmission using dynamic antenna index assignment," {\it IEEE GLOBECOM 2017}, Singapore, 2017, Dec. 2017, pp. 1-6.
	
	\bibitem{Wei} Y. Wei, L. Wang and T. Svensson, \textquotedblleft Analysis of secrecy rate against eavesdroppers in MIMO modulation systems," \textit{2015 Int. Conf. Wireless Commun. Signal Process. (WCSP)}, Nanjing, Oct. 2015, pp. 1-5.
	
	
	
	
	%
	
	
	%
	
	\bibitem{parallel} Z. Li, R. D. Yates, and W. Trappe, \textquotedblleft Secrecy capacity of independent parallel channels,” in \textit{44th Annual Allerton CCC}, Sep. 2006.
	
	\bibitem{Renna_cap} F. Renna, N. Laurenti, and H. V. Poor, \textquotedblleft Physical-layer secrecy for OFDM transmissions over fading channels,” \textit{IEEE Trans. Inf. Forens. Security}, vol. 7, no. 4, pp. 1354–1367, Aug. 2012.
	
	\bibitem{Renna}
	F. Renna, N. Laurenti, and H. V. Poor, \textquotedblleft Achievable secrecy rates for wiretap OFDM with QAM constellations,'' in \textit{Proc. 5th Int. ICST Conf. Perform. Eval. Method. VALUETOOLS}, Paris, France, 2011,
	pp. 679–-686.
	
	\bibitem{Yusuf} M. Yusuf and H. Arslan, \textquotedblleft Controlled inter-carrier interference for physical layer security in OFDM systems," \textit{2016 IEEE  Veh. Technol. Conf. (VTC Fall)}, Montreal, QC, 2016, pp. 1-5.
	
	\bibitem{HaoLi} H. Li, X. Wang and J. Y. Chouinard, "Eavesdropping-resilient OFDM system using sorted subcarrier interleaving," in \textit{IEEE Trans. Wireless Commun.}, vol. 14, no. 2, pp. 1155-1165, Feb. 2015.
	
	
	\bibitem{HQin}
	H. Qin, Y. Sun, T. Chang, X. Chen,; C. Chi, M. Zhao, J. Wang, \textquotedblleft Power allocation and time-domain artificial noise design
	for wiretap OFDM with discrete inputs," \textit{IEEE Trans. Wireless Commun.},
	vol. 12, no. 6, pp. 2717–-2729, Jun. 2013.
	
	\bibitem{Akitaya}
	T. Akitaya, S. Asano, and T. Saba, \textquotedblleft Time-domain artificial noise generation
	technique using time-domain and frequency-domain processing
	for physical layer security in MIMO-OFDM systems," in \textit{Proc. IEEE Int. Conf. Commun. Workshops (ICCW)}, Jun. 2014, pp. 807-–812.

	\bibitem{Kwan2012}
    D. W. K. Ng, E. S. Lo and R. Schober, \textquotedblleft Energy-efficient resource allocation for secure OFDMA systems," in \textit{IEEE Trans. on Vehic. Technol.}, vol. 61, no. 6, pp. 2572-2585, July 2012.

	\bibitem{Li2013}
	H. Li, X. Wang and W. Hou, \textquotedblleft Secure transmission in OFDM systems by using time domain scrambling," \textit{2013 IEEE 77th Vehic. Technol. Conf. (VTC Spring)}, Dresden, 2013, pp. 1-5.
	
	\bibitem{HAMAMREH2017B}
	J. M. Hamamreh, E. Basar and H. Arslan, \textquotedblleft OFDM-subcarrier index selection for enhancing security and reliability of 5G URLLC services," in \textit{IEEE Access}, vol. 5, pp. 25863-25875, 2017.
	
	\bibitem{Hamamreh2017}
	J. M. Hamamreh, H. M. Furqan and H. Arslan, \textquotedblleft Secure pre-coding and post-coding for OFDM systems along with hardware implementation," \textit{2017 13th International Wireless Communications and Mobile Computing Conference (IWCMC)}, Valencia, 2017, pp. 1338-1343.
	
	\bibitem{Hajomer2018}
	A. A. E. Hajomer, X. Yang and W. Hu, \textquotedblleft Secure OFDM transmission precoded by chaotic discrete Hartley transform," in \textit{IEEE Photonics Journal}, vol. 10, no. 2, pp. 1-9, April 2018.
	
	\bibitem{Yusuff} M. Yusuf, H. Arslan, \textquotedblleft On signal space diversity: An adaptive interleaver for enhancing physical layer security in frequency selective fading channels," \textit{Phys. Commun.}, vol. 24,
	pp. 154-160, Sep. 2017.
	
	\bibitem{Furkan2017}
	H. M. Furqan, J. M. Hamamreh and H. Arslan, \textquotedblleft Enhancing physical layer security of OFDM systems using channel shortening," \textit{IEEE Int. Symp. Pers. Indoor Mobile Radio Commun. (PIMRC 2017)}, Montreal, QC, 2017, pp. 1-5.
	
	\bibitem{Farhang2016}
	B. Farhang-Boroujeny and H. Moradi, \textquotedblleft OFDM Inspired Waveforms for 5G," in \textit{IEEE Commun. Surveys Tuts.}, vol. 18, no. 4, pp. 2474-2492, Fourth Quarter 2016.
	
	\bibitem{Schellmann2014}
	M. Schellmann et al., \textquotedblleft FBMC-based air interface for 5G mobile: challenges and proposed solutions," \textit{2014 9th Int. Conf. on CROWNCOM}, Oulu, 2014, pp. 102-107.
	
	\bibitem{SSBook}
	D. Torrieri, \textit{Principles of Spread-Spectrum Communication Systems}, 2nd
	edition. Springer, 2011.
	
		\bibitem{CDMABook}
	M. A. Abu-Rgheff, \textit{Introduction to CDMA Wireless Communications},
	1st ed. London, U.K., Academic Press, 2007.
		
	\bibitem{Li2005}
	Tontong Li, Jian Ren, Qi Ling and Anil Jain, \textquotedblleft Physical layer built-in security analysis and enhancement of CDMA systems," \textit{IEEE MILCOM 2005}, Atlantic City, NJ, 2005, pp. 956-962 Vol. 2.
	
	
	\bibitem{Mheich} Z. Mheich, F. Alberge, and P. Duhamel, \textquotedblleft Achievable secrecy rates for the broadcast channel with confidential message and finite constellation inputs," \textit{IEEE Trans. Commun.}, vol. 63, no. 1, pp. 195--205, Jan. 2015.
	
	\bibitem{MAC_Allen} T. Allen, A. Tajer and N. Al-Dhahir, \textquotedblleft Secure Alamouti MAC transmissions," in \textit{IEEE Trans. Wireless Commun.}, vol. 16, no. 6, pp. 3674-3687, Jun. 2017.
	
	
	
	\bibitem{Jin16} J. Jin, C. Xiao, M. Tao and W. Chen, \textquotedblleft Linear precoding for cognitive multiple access wiretap channel with finite-alphabet inputs," 2016 \textit{IEEE ICC}, Kuala Lumpur, May 2016, pp. 1-6.
	
	\bibitem{Zeng16} W. Zeng, Y. R. Zheng and C. Xiao, \textquotedblleft Multiantenna secure cognitive radio networks with finite-alphabet inputs: a global optimization approach for precoder design," in \textit{IEEE Trans. Wireless Commun.}, vol. 15, no. 4, pp. 3044-3057, Apr. 2016.
	
	\bibitem{Cao18} K. Cao, Y. Cai, Y. Wu, W. Yang and X. Guan, \textquotedblleft Secure communication for MISO secrecy channel with multiple multiantenna eavesdroppers having finite alphabet inputs," in \textit{IEEE Access}, vol. 6, no. 1, pp. 7402-7411.
	
	\bibitem{Vishwakarma14} S. Vishwakarma and A. Chockalingam, \textquotedblleft Power allocation in MIMO wiretap channel with statistical CSI and finite-alphabet input," \textit{2014 SPCOM}, Bangalore, Jul. 2014, pp. 1-6.
	
	
	\bibitem{Vishwakarma} S. Vishwakarma and A. Chockalingam, \textquotedblleft Decode-and-forward relay beamforming for secrecy with finite-alphabet input," in \textit{IEEE Commun. Lett.}, vol. 17, no. 5, pp. 912-915, May 2013.
	
	\bibitem{Cao17} K. Cao, Y. Cai, Y. Wu and W. Yang, \textquotedblleft Cooperative jamming for secure communication with finite alphabet inputs," in \textit{IEEE Commun. Lett.}, vol. 21, no. 9, pp. 2025-2028, Sep. 2017.
	
	\bibitem{Cao_Helper} K. Cao, Y. Wu, Y. Cai and W. Yang, \textquotedblleft Secure transmission with aid of a helper for MIMOME network having finite alphabet inputs," in IEEE Access, vol. 5, pp. 3698-3708, Feb. 2017.
	
	\bibitem{Furqan_SW} H. M. Furqan, J. M. Hamamreh and H. Arslan, \textquotedblleft Secure communication via untrusted switchable decode-and-forward relay," \textit{2017 13th International Wireless Communications and Mobile Computing Conference (IWCMC)}, Valencia, 2017, pp. 1333-1337.
	
	\bibitem{Larsson} E. Larsson et al., \textquotedblleft Massive MIMO for next generation wireless systems,” \textit{IEEE Commun. Mag.}, vol. 52, no. 2, Feb. 2014, pp. 186–95.
	
	\bibitem{Rappaport} T. Rappaport et al., \textquotedblleft Millimeter wave mobile communications for 5G cellular: It will work!," \textit{IEEE Access}, vol. 1, May 2013, pp. 335–49.
	
	
	\bibitem{JWang} J. Wang, J. Lee, F. Wang, and T. Q. S. Quek, \textquotedblleft Jamming-aided secure communication in massive MIMO Rician channels,” \textit{IEEE Trans. Commun.}, vol. 64, pp. 6854–6868, Dec. 2015.
	
	\bibitem{KGou} K. Guo, Y. Guo, and G. Ascheid, “Security-constrained power allocation in mu-massive-MIMO with distributed antennas,” \textit{IEEE Trans. Wireless Commun.}, vol. 15, pp. 8139–8153, Dec. 2016.
	
	\bibitem{JZhu} J. Zhu, R. Schober, and V. K. Bhargava, \textquotedblleft Linear precoding of data and artificial noise in secure massive MIMO systems,” \textit{IEEE Trans. Wireless Commun.}, vol. 15, pp. 2245–2261, Mar. 2016.
	
	\bibitem{Vuppala} S. Vuppala, S. Biswas, and T. Ratnarajah, \textquotedblleft An analysis on secure communication in millimeter/micro-wave hybrid networks,” \textit{IEEE Trans. Commun.}, vol. 64, pp. 3507–3519, Aug. 2016.
	
	\bibitem{CWang} C. Wang and H.-M. Wang, \textquotedblleft Physical layer security in millimeter wave cellular networks,” \textit{IEEE Trans. Wireless Commun.}, vol. 15, pp. 5569–5585, Aug. 2016.
	
	\bibitem{YZhu} Y. Zhu, L. Wang, K. K. Wong and R. W. Heath, \textquotedblleft Secure communications in
	millimeter wave ad hoc networks,"  \textit{IEEE Trans. Wireless Commun.}, vol. 16, no. 5, pp. 3205-3217, May 2017.
	
	\bibitem{ZZhang} Z. Zhang, X. Chai, K. Long, A. V. Vasilakos and L. Hanzo, \textquotedblleft Full duplex techniques for 5G networks: self-interference cancellation, protocol design, and relay selection," in \textit{IEEE Commun. Mag.}, vol. 53, no. 5, pp. 128-137, May 2015.
	
	\bibitem{LDai} L. Dai, B. Wang, Y. Yuan, S. Han, C. l. I and Z. Wang, \textquotedblleft Non-orthogonal multiple access for 5G: solutions, challenges, opportunities, and future research trends," in \textit{IEEE Commun. Mag.}, vol. 53, no. 9, pp. 74-81, Sep. 2015.
	
	\bibitem{Valli13} N. Valliappan, A. Lozano and R. W. Heath, \textquotedblleft Antenna subset modulation for secure millimeter-wave wireless communication," in \textit{IEEE Trans. Commun.}, vol. 61, no. 8, pp. 3231-3245, Aug. 2013.
	
	\bibitem{JFan17} J. Fan and M. Wu, \textquotedblleft Antenna selection in switched phased array architecture for secure millimeter wave communication," 2017 IEEE 17th Int. Conf. on Commun. Techn. (ICCT), Chengdu, 2017, pp. 718-723.
	
	\bibitem{YLiu17} Y. Liu, Z. Qin, M. Elkashlan, Y. Gao and L. Hanzo, \textquotedblleft Enhancing the physical layer security of non-orthogonal multiple access in large-scale networks," in \textit{IEEE Wireless Commun.}, vol. 16, no. 3, pp. 1656-1672, Mar. 2017.
	
	\bibitem{DXu} D. Xu, P. Ren, Q. Du, L. Sun and Y. Wang, \textquotedblleft Design for NOMA: combat eavesdropping and improve spectral efficiency in the two-user relay network," \textit{Proc. Global Commun. Conf. (Globecom 2017)}, Singapore, 2017, pp. 1-6.
	
	\bibitem{QLi16} Q. Li, W.-K. Ma, and D. Han, \textquotedblleft Sum secrecy rate maximization for full-duplex two-way relay networks using Alamouti-based rank-two beamforming,” \textit{IEEE J. Sel. Topic Signal Process.}, vol. 10, pp. 1359-1374, Dec. 2016.
	
	\bibitem{HamamrehOTDM} J. M. Hamamreh and H. Arslan, \textquotedblleft Secure orthogonal transform division multiplexing (OTDM) waveform for 5G and beyond," in \textit{IEEE Commun. Lett.}, vol. 21, no. 5, pp. 1191-1194, May 2017.
	
	\bibitem{IoT} A. Mukherjee, \textquotedblleft Physical-layer security in the Internet of things: sensing and communication confidentiality under resource constraints," \textit{Proceedings of IEEE}, vol. 103, no. 10, pp. 1747-1761, Oct. 2015.
	

	
	\bibitem{survey}
	M. R. Bloch, M. Hayashi, and A. Thangaraj, \textquotedblleft Error-control coding for physical-layer secrecy," \textit{Proceedings of IEEE}, vol. 103, no. 10, pp. 1725--1746, Oct. 2015.
	
	\bibitem{ozarow}
	L. H. Ozarow and A. D. Wyner, \textquotedblleft Wire-tap channel II," \textit{Bell Lab. Tech. J.}, vol. 63, no. 10, pp. 2135-2157, Dec. 1984.
	
	\bibitem{syndrome}
	G. Cohen and G. Zemor, \textquotedblleft Syndrome-coding for the wiretap channel revisited," in \textit{Proc. IEEE Information Theory Workshop}, Chengdu, China, Oct. 2006, pp. 33–36.
	
	
	\bibitem{lattice}
	F. Oggier, P. Sole and J. Belfiore, \textquotedblleft Lattice codes for the wiretap Gaussian channel: construction and analysis," \textit{IEEE Trans. Inf. Theory}, vol. 62, no. 10, pp. 5690-5707, Oct. 2016.
	
	\bibitem{lattice2}
	C. Ling, L. Luzzi, J.-C. Belfiore, and D. Stehlé, \textquotedblleft Semantically secure lattice
	codes for the Gaussian wiretap channel," \textit{Computing Research Repository}, Oct. 2012, pp. 1–19.
	
	\bibitem{puncture}
	D. Klinc, J. Ha, S. M. McLaughlin, J. Barros, and B.-J. Kwak, \textquotedblleft LDPC codes for the Gaussian wiretap channel," \textit{IEEE Trans. Inf. Forensics Security}, vol. 6, no. 3, pp. 532-540, Sep. 2011.
	
	\bibitem{BERbased}
	I. M. Kim, B. H. Kim, and J. K. Ahn, \textquotedblleft BER-based physical layer security with finite code length: Combining strong converse and error amplification," \textit{IEEE Trans. Commun.}, vol. 64, pp. 3844-3857, Sep. 2016.
	
	\bibitem{scramble}
	M. Baldi, M. Bianchi, and F. Chiaraluce, \textquotedblleft Non-systematic codes for
	physical layer security," in \textit{Proc. IEEE Information Theory Workshop (ITW 2010)}, Dublin, Ireland, Aug. 2010.
	
	\bibitem{scramblesecond}
	M. Baldi, F. Bambozzi, and F. Chiaraluce, \textquotedblleft Coding with scrambling, concatenation, and HARQ for the AWGN wire-tap channel: A security gap analysis," \textit{IEEE Trans. Inf. Forensics Security}, vol. 7, no. 3, pp. 883–-894, Jun. 2012.
	
\bibitem{RandomizedConvolutional}
A. Nooraiepour and T. M. Duman, \textquotedblleft Randomized convolutional codes for the wiretap channel," in \textit{IEEE Trans. Commun.}, vol. 65, no. 8, pp. 3442-3452, Aug. 2017.
	
	
\bibitem{polar-LDPC}
Y. X. Zhang, A. Liu, \textquotedblleft Polar-LDPC concatenated coding for the AWGN wiretap channel," \textit{IEEE Commun. Lett.}, vol. 18, no. 10, pp. 1683-1686, Oct. 2014.
	
\bibitem{RandomizedTurbo}
A. Nooraiepour and T. M. Duman \textquotedblleft Randomized turbo codes for the wiretap channel," in \textit{IEEE GLOBECOM}, Singapore, Dec. 2017.

\bibitem{RandomizedSCLDGM}
A. Nooraiepour and T. M. Duman, \textquotedblleft Randomized serially concatenated LDGM codes for the Gaussian wiretap channel," Accepted for publication in \textit{IEEE Commun. Lett.}, Dec. 2017.

\bibitem{BALDI2017}
M. Baldi, L. Senigagliesi and F. Chiaraluce, \textquotedblleft On the security of transmissions over fading wiretap channels in realistic conditions," \textit{2017 IEEE Int. Conf. on Commun. (ICC)}, Paris, 2017, pp. 1-6.
	
\bibitem{lowerbound1}
C. W. Wong, T. F. Wong, and J. M. Shea, \textquotedblleft LDPC code design for the BPSK-constrained Gaussian wiretap channel," in \textit{Proc. IEEE GLOBECOM 2011 Workshops}, Houston, TX, Dec. 2011, pp. 898–902.
	
\bibitem{lowerbound2}
	M. Baldi, G. Ricciutelli, N. Maturo, and F. Chiaraluce, \textquotedblleft Performance assessment and design of finite length LDPC codes for the gaussian wiretap channel," \textit{IEEE ICC 2015 Workshop on Physical-Layer Security}, Jun. 2015, pp. 435-440.
	
\bibitem{explicit}
	H. Tyagi and A. Vardy, \textquotedblleft Explicit capacity-achieving coding scheme for the Gaussian wire- tap channel," in \textit{IEEE Int. Symp. Inf. Theory}, Jul. 2014, pp. 956–-960.
	
\bibitem{finite}
	W. Yang, R. F. Schaefer, and H. V. Poor, ``Finite-blocklength bounds for wiretap channels,'' in \textit{IEEE Int. Symp. Inf. Theory}, Barcelona, Spain, pp. 3087–-3091, Jul. 2016.
	
\bibitem{Zheng} M. Zheng, M. Tao, and W. Chen., \textquotedblleft Polar coding for secure
transmission in MISO fading wiretap channels." [Online]. Available:
http://arxiv.org/abs/1411.2463, Nov. 2014.

\bibitem{Si} H. Si, O. O. Koyluoglu and S. Vishwanath, \textquotedblleft Hierarchical Polar Coding for Achieving Secrecy Over State-Dependent Wiretap Channels Without Any Instantaneous CSI," in \textit{IEEE Trans. Commun.}, vol. 64, no. 9, pp. 3609-3623, Sep. 2016.	
	
	
\bibitem{Luzzi} L. Luzzi, C. Ling and R. Vehkalahti, \textquotedblleft Almost universal codes for fading wiretap channels," \textit{IEEE Int. Symp. Inf. Theory (ISIT)}, Barcelona, Jul. 2016, pp. 3082-3086.	


	
	
	
	%
	%
	%
	%
	%
	%
	%
	%
	%
	%
	%
	%
	%
	%
	%
	%
	%
	
	\bibitem{Shamai_Verdu} S. Shamai and S. Verdu, \textquotedblleft Worst-case power-constrained noise for binary input channels," \textit{IEEE Trans. Inf. Theory}, vol. 38, no. 5, pp. 1494- 1511, Sep. 1992.
	
	\bibitem{Ayca} A. Ozcelikkale and T. M. Duman, \textquotedblleft Cooperative precoding and artificial noise design for security over interference channels," \textit{IEEE Signal Process. Lett.}, vol. 22, pp. 2234-2238, Dec. 2015.
	
	\bibitem{Chen} X. Chen, D. W. K. Ng, and H.-H. Chen, \textquotedblleft Secrecy wireless information and power transfer: challenges and opportunities," \textit{IEEE Commun. Mag.}, May 2016.
	
	\bibitem{R_Zhang13} R. Zhang and C. K. Ho, \textquotedblleft MIMO broadcasting for simultaneous wireless information and power transfer," \textit{IEEE Trans. Wireless Commun.}, vol. 12, no. 5, pp. 1989-–2001, May 2013.
	
	\bibitem{Zhang16} J. Zhang, C. Yuen, C. K. Wen, S. Jin, K. K. Wong and H. Zhu, \textquotedblleft Large system secrecy rate analysis for SWIPT MIMO wiretap channels," \textit{IEEE Trans. Inf. Forensics Security}, vol. 11, no. 1, pp. 74--85, Jan. 2016.
	
	\bibitem{Ulukus16} K. Banawan and S. Ulukus, \textquotedblleft MIMO wiretap channel under receiver side power constraints with applications to wireless information transfer and cognitive radio,” \textit{IEEE Trans. on Commun.}, vol. 64, no. 9, pp. 3872--3885, Sep. 2016.
	
	
	\bibitem{Wang} H. M. Wang, T. X. Zheng, J. Yuan, D. Towsley and M. H. Lee, \textquotedblleft Physical layer security in heterogeneous cellular networks," in \textit{IEEE Trans. Commun.}, vol. 64, no. 3, pp. 1204-1219, Mar. 2016.
	
	\bibitem{YZhou} Y. Zhou, Z. Z. Xiang, Y. Zhu, and Z. Xue, \textquotedblleft Application of full-duplex wireless technique into secure MIMO communication: Achievable secrecy rate based optimization,” \textit{IEEE Signal Process. Lett.}, vol. 21, pp. 804–808, Jul. 2014.
	
	\bibitem{Abedi} M. R. Abedi, N. Mokari, H. Saeedi, \textquotedblleft How to manage resources to provide physical layer security: Active versus passive adversary?," \textit{Phys. Commun.}, vol. 27, pp. 143-149, Apr. 2018.

	\bibitem{Ogg} J. Lu, J. Harshan, F. Oggier,
	\textquotedblleft Performance of lattice coset codes on universal software radio peripherals," \textit{Phys. Commun.}, vol. 24, pp. 66-72, Apr. 2017.
	
	\bibitem{Ham_cross} J. M. Hamamreh, M. Yusuf, T. Baykas and H. Arslan, \textquotedblleft Cross MAC/PHY layer security design using ARQ with MRC and adaptive modulation," \textit{2016 IEEE Wireless Communications and Networking Conference}, Doha, 2016, pp. 1-7.
	
    \bibitem{LZhou14} L. Zhou, D. Wu, B. Zheng, and M. Guizani, \textquotedblleft Joint physical-application layer security for wireless multimedia delivery," \textit{IEEE Commun. Mag.}, vol. 52, pp. 66-72, Mar. 2014.
	
	
\end{thebibliography}
\end{document}